\documentclass[twocolumn,superscriptaddress,amsmath,amssymb,aps,pra,floatfix]{revtex4-2}
\usepackage[utf8]{inputenc}
\usepackage[english]{babel}
\usepackage{wrapfig}
\usepackage{float}
\usepackage{amssymb}
\usepackage{amsfonts}
\usepackage{amsmath}
\usepackage{cases}
\usepackage{stmaryrd}
\usepackage{tipa}
\usepackage[dvips]{graphicx}
\usepackage{dcolumn}
\usepackage{bbold}
\usepackage{graphicx}
\usepackage{graphicx}
\usepackage{subfigure}
\usepackage{dcolumn}
\usepackage{bm}
\usepackage{color}
\usepackage{xcolor}
\usepackage{CJK}
\usepackage{subfigure}
\usepackage{physics}
\usepackage{array}
\usepackage{multirow}
\usepackage{mathtools}
\usepackage{enumerate}
\usepackage{comment}

\usepackage[T1]{fontenc}
\usepackage{verbatim}
\usepackage{xcolor}
\usepackage{pgf,tikz}
\usepackage{mathrsfs}

\usepackage{amsmath,amssymb,textcomp}
\everymath{\displaystyle}

\usepackage{times}
\usepackage{tgheros}

\usepackage[english]{babel}
\usepackage{eurosym}
\usepackage{siunitx}

\usepackage{graphicx}
\graphicspath{{./img/}}
\usepackage{svg}
\usepackage[section]{placeins}

\usepackage{hyperref}
\usepackage[normalem]{ulem}

\begin{document}
 
\author{F.\,Fesquet}
\email[]{florian.fesquet@wmi.badw.de}
\affiliation{Walther-Mei{\ss}ner-Institut, Bayerische Akademie der Wissenschaften, 85748 Garching, Germany}
\affiliation{Physik-Department, Technische Universit\"{a}t M\"{u}nchen, 85748 Garching, Germany}

\author{F.\,Kronowetter}
\affiliation{Walther-Mei{\ss}ner-Institut, Bayerische Akademie der Wissenschaften, 85748 Garching, Germany}
\affiliation{Physik-Department, Technische Universit\"{a}t M\"{u}nchen, 85748 Garching, Germany}
\affiliation{Rohde \& Schwarz GmbH \& Co. KG, Mühldorfstraße 15, 81671 Munich, Germany}

\author{M.\,Renger}
\affiliation{Walther-Mei{\ss}ner-Institut, Bayerische Akademie der Wissenschaften, 85748 Garching, Germany}
\affiliation{Physik-Department, Technische Universit\"{a}t M\"{u}nchen, 85748 Garching, Germany}

\author{Q.\,Chen}
\affiliation{Walther-Mei{\ss}ner-Institut, Bayerische Akademie der Wissenschaften, 85748 Garching, Germany}
\affiliation{Physik-Department, Technische Universit\"{a}t M\"{u}nchen, 85748 Garching, Germany}

\author{K.\,Honasoge}
\affiliation{Walther-Mei{\ss}ner-Institut, Bayerische Akademie der Wissenschaften, 85748 Garching, Germany}
\affiliation{Physik-Department, Technische Universit\"{a}t M\"{u}nchen, 85748 Garching, Germany}

\author{O.\,Gargiulo}
\affiliation{Walther-Mei{\ss}ner-Institut, Bayerische Akademie der Wissenschaften, 85748 Garching, Germany}
\affiliation{Physik-Department, Technische Universit\"{a}t M\"{u}nchen, 85748 Garching, Germany}

\author{Y.\,Nojiri}
\affiliation{Walther-Mei{\ss}ner-Institut, Bayerische Akademie der Wissenschaften, 85748 Garching, Germany}
\affiliation{Physik-Department, Technische Universit\"{a}t M\"{u}nchen, 85748 Garching, Germany}

\author{A.\,Marx}
\affiliation{Walther-Mei{\ss}ner-Institut, Bayerische Akademie der Wissenschaften, 85748 Garching, Germany}

\author{F.\,Deppe}
\affiliation{Walther-Mei{\ss}ner-Institut, Bayerische Akademie der Wissenschaften, 85748 Garching, Germany}
\affiliation{Physik-Department, Technische Universit\"{a}t M\"{u}nchen, 85748 Garching, Germany}
\affiliation{Munich Center for Quantum Science and Technology (MCQST), Schellingstr. 4, 80799 Munich, Germany}

\author{R.\,Gross}
\affiliation{Walther-Mei{\ss}ner-Institut, Bayerische Akademie der Wissenschaften, 85748 Garching, Germany}
\affiliation{Physik-Department, Technische Universit\"{a}t M\"{u}nchen, 85748 Garching, Germany}
\affiliation{Munich Center for Quantum Science and Technology (MCQST), Schellingstr. 4, 80799 Munich, Germany}

\author{K.\,G.\,Fedorov}
\email[]{kirill.fedorov@wmi.badw.de}
\affiliation{Walther-Mei{\ss}ner-Institut, Bayerische Akademie der Wissenschaften, 85748 Garching, Germany}
\affiliation{Physik-Department, Technische Universit\"{a}t M\"{u}nchen, 85748 Garching, Germany}

\title{Perspectives of microwave quantum key distribution in open-air} 

\begin{abstract}
One of the cornerstones of quantum communication is the unconditionally secure distribution of classical keys between remote parties. This key feature of quantum technology is based on the quantum properties of propagating electromagnetic waves, such as entanglement, or the no-cloning theorem. However, these quantum resources are known to be susceptible to noise and losses, which are omnipresent in open-air communication scenarios. In this work, we theoretically investigate the perspectives of continuous-variable open-air quantum key distribution at microwave frequencies. In particular, we present a model describing the coupling of propagating microwaves with a noisy environment. Using a protocol based on displaced squeezed states, we demonstrate that continuous-variable quantum key distribution with propagating microwaves can be unconditionally secure at room temperature up to distances of around 200 meters. Moreover, we show that microwaves can potentially outperform conventional quantum key distribution at telecom wavelength at imperfect weather conditions.
\end{abstract}

\maketitle

\section{Introduction} 

Quantum key distribution (QKD) can be defined as a method to securely exchange a common key between two parties, conventionally referred to as Alice and Bob. QKD has attracted increasing interest over the last decades due to the promise of unconditionally secure communication while maintaining high secret key rates \cite{Comandar2014,Wang2018}. The security of commonly used classical encryption schemes, or key agreement protocols, is based on asymmetric mathematical problems which cannot be easily inverted by classical methods. Prominent examples are the Diffie-Hellman algorithm or the RSA code \cite{DiffieHellman1976,RSA1978}. In contrast, QKD relies on the fact that unconditional security is provided by the fundamental laws of quantum mechanics, taking advantage of unique quantum resources, such as entanglement \cite{Scarani2009}, or notably, the no-cloning theorem \cite{Zurek1973}. In QKD based on continuous-variable (CV) quantum states (CV-QKD), one encodes information in conjugate electromagnetic field quadratures, according to various specific protocols \cite{Laudenbach2018}. In particular, CV-QKD protocols provide a powerful alternative to QKD based on discrete-variable (DV) quantum states (DV-QKD), due to potentially higher secret key rates and being compatible with the existing communication infrastructure \cite{Lodewyck2005,Grosshans2003}.

First successful realizations of QKD protocols have been implemented at the near-infrared regime at so-called telecom wavelengths (780-850\,\si{\nano\meter} and 1520-1600\,\si{\nano\meter} wavelength) \cite{Buttler2000}, where a significant progress has been achieved with the recent realization of ground-to-satellite QKD networks \cite{Liao2017}. A particular reason for this choice of frequencies is the low atmospheric absorption on the order of \SI{1.0e-2}{\deci\bel/\kilo\meter} \cite{Kaushal2018}. However, in the light of recent impressive achievements with superconducting quantum circuits operating in the microwave regime \cite{Arute2019, Kjaergaard2020, Pogorzalek2019, Bienfait2019} and due to the lack of efficient microwave-to-optical frequency transducers, a natural choice is to consider quantum communication and QKD in the associated microwave regime. Here, superconducting Josephson parametric amplifiers (JPAs) represent a robust source of quantum states in the form of squeezed microwaves. Flux-driven JPAs routinely generate microwave squeezed states with squeezing levels up to \SI{10}{\deci\bel}  below the vacuum limit \cite{Zhong2013,Eichler2014,Grimsmo2017}. 

In this work, we focus on a theoretical analysis of a particular one-way communication prepare-and-measure CV-QKD protocol with Gaussian modulation \cite{Cerf2001} for the microwave wavelength range of 30-300\,\si{\milli\meter}, corresponding to the frequency range of  1-10\,\si{\giga\hertz}. We analyze the potential of this protocol for open-air microwave quantum communication (MQC),  considering realistic atmospheric conditions. We compare its performance to a traditional implementation at the telecom wavelength of \SI{1550}{\nano\meter} $\left(\SI{193.55}{\tera\hertz}\right)$. We model the signal readout with a homodyne detection, since the information is encoded only in one of the two electromagnetic field quadratures. In the last step, signal detection is followed by a one-way classical reconciliation, or error correction, protocol. There, an important distinction must be made between either direct reconciliation (DR) or reverse reconciliation (RR) \cite{grosshans2003_3}. In DR, the one-way communication is performed from Alice to Bob. As a result, Alice's key is used as a re\-ference which Bob tries to estimate from the data he obtained after the communication. On contrary in RR, the one-way communication goes from Bob to Alice, where Bob's measured key is estimated by Alice. 

In Sections \ref{subsec:Secret_key_and_secret_key_rate} and \ref{subsec:Experimental_setup}, we introduce all relevant elements for a practical implementation of the CV-QKD protocol in the microwave regime. We perform a quantitative analysis of the secret key rate and secure distance, assuming either DR or RR with an ideal reconciliation efficiency in the asymptotic case, i.e., in the case of Alice and Bob exchanging an infinitely long key. We additionally present a model based on beam splitter transformations to describe the coupling of propagating microwaves with a bright microwave thermal background. Our results show that microwave CV-QKD can be well suited for short-distance open-air communication scenarios. In particular, in Section \ref{subsec:Security} we find that the microwave CV-QKD protocol may produce higher secure key rates than its telecom counterpart. We conclude our discussion in Section \ref{sec:weather_imperfections} with a study of weather effects for the particular common cases of rain and haze. We find that MQC secure communication distances are almost unchanged as compared to ideal dry weather conditions. This is in striking contrast to the telecom protocols.

\begin{figure}[t]
	\centering
	\includegraphics[width=0.8\columnwidth]{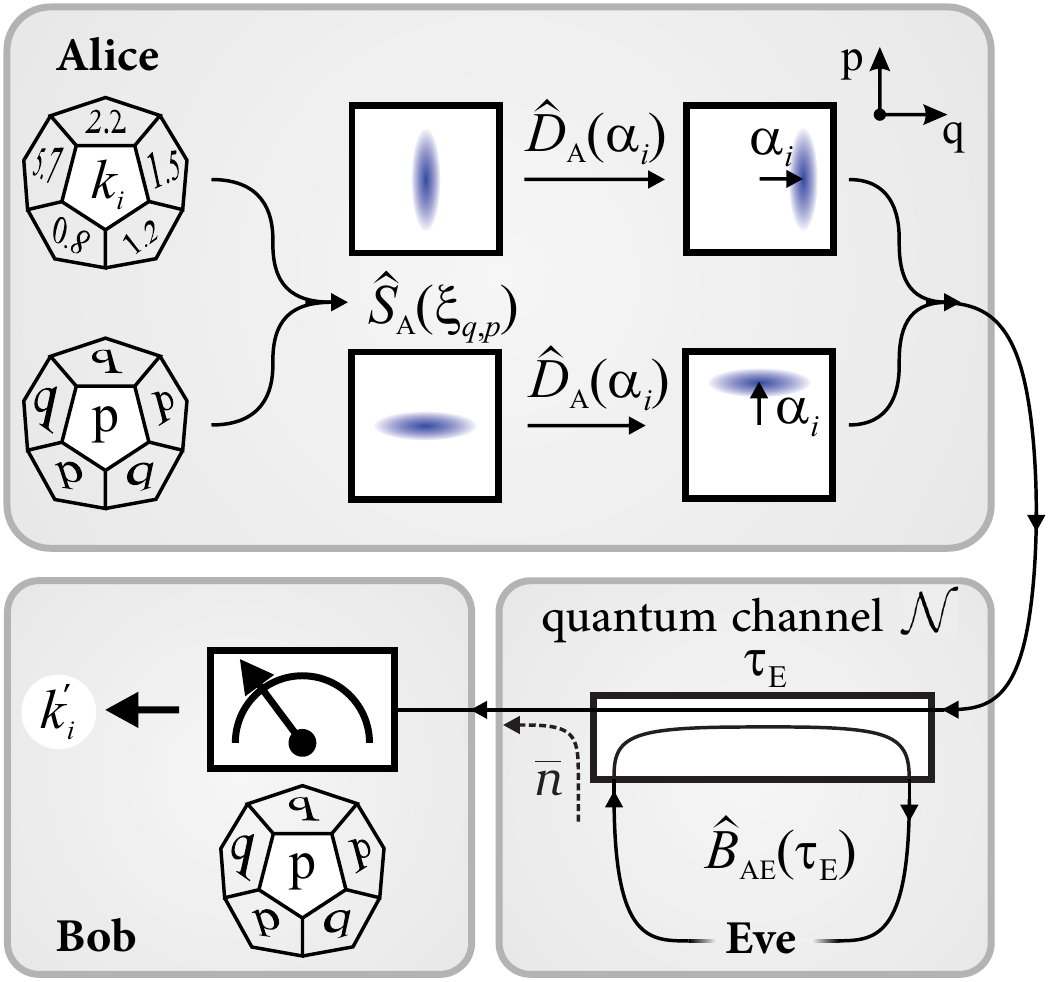}
	\caption{General scheme of a QKD protocol based on displaced squeezed states. Here, Alice starts by generating a random continuous-variable number $k_{i}$ corresponding to Alice's symbol and randomly choosing among one of the two possible encoding bases, $q$ or $p$. This procedure results in propagating states which are squeezed along one or the other quadrature with the complex squeezing amplitude $\xi$. Every symbol $k_{i}$ is encoded via a displacement amplitude $\alpha_i$, such that $|\alpha_{i}| = |k_{i}|$. The resulting displaced squeezed state is sent through a quantum channel $\mathcal{N}$ with transmissivity $\tau_{\mathrm{E}}$. This channel is assumed to be under full control of a potential eavesdropper, Eve, who also induces extra noise photons $\bar{n}$. At the end, Bob receives a state which he measures in a random basis $q$ or $p$ in order to obtain an estimate, $k_{i}^{\prime}$, of the original symbol.}
	\label{fig:Fig_1}
\end{figure}

\section{Quantum key distribution protocol with continuous variables}
\label{subsec:Secret_key_and_secret_key_rate}

First, we consider a prepare-and-measure CV-QKD protocol, independent of a particular hardware platform and frequency range, as described in Ref.\,\citenum{Cerf2001}. A corresponding scheme is shown in Fig.\,\ref{fig:Fig_1}. Here, Alice transmits to Bob a Gaussian-modulated random key $\mathcal{K} = \{k_{\mathrm{1}}, ..., k_{i}, ..., k_{\mathrm{n}}\} $. This key is a string of real numbers $k_{i}$, randomly chosen from a Gaussian distribution with variance $\sigma_{\mathrm{A}}^{2}$. To this end, Alice prepares a $q$-squeezed or $p$-squeezed state \cite{Pogorzalek2019}, with both states having an equal chance of being selected. Each symbol, $k_{i}$, is then encoded as a displacement amplitude, $\alpha_i$, of each squeezed state such that $|\alpha_i| = |k_{i}|$. Averaging over various states of Alice results in a thermal state, preventing Eve from extracting any information on the encoding basis and encoded symbols. This imposes an additional cons\-traint $\sigma_{\mathrm{s}}^{2} + \sigma_{\mathrm{A}}^{2} = \sigma_{\mathrm{as}}^{2}$, where we denote the squeezed and antisqueezed quadrature variance as $\sigma_{\mathrm{s}}^{2}$ and $\sigma_{\mathrm{as}}^{2}$, respectively. The resulting displaced squeezed state propagates through a lossy and noisy quantum channel $\mathcal{N}$ before being received and measured by Bob with a local homodyne measurement. After repeating this process for each of Alice's symbols, Bob obtains a measured key $\mathcal{K^{\prime}} = \{k^{\prime}_{\mathrm{1}}, ..., k^{\prime}_{i}, ..., k^{\prime}_{\mathrm{n}}\} $, representing an estimation of Alice's key $\mathcal{K}$. 

In order to obtain a common secret key, Alice and Bob perform a one-way classical postprocessing. The first step, know as sifting of the keys, consists of producing compatible data, discarding any measurement where encoding and measurement basis disagree. The second step, commonly referred to as parameter estimation, allows to obtain an upper bound on the amount of information lost in the quantum channel. Then, a classical reconciliation algorithm, depending on whether DR or RR has been chosen, is used to generate a common key. This performance of this algorithm is characterized with a reconciliation efficiency $\beta$. Alice and Bob further proceed to a confirmation step to validate a recovered common key. Finally, a classical privacy amplification algorithm produces a secret key, discarding any eavesdropped bits common key.

To describe the quantum channel $\mathcal{N}$, we quantify losses using a quantum channel transmissivity $\tau_{\mathrm{E}}$ and quantum channel noise, $\bar{n}$ which represents an average noise photon number referred to the output of the channel. Generally, these losses and noise represent interactions between the propagating states and environment. In the worst case, from the standpoint of security, the quantum channel is under the full control of a potential eavesdropper, Eve. This implies that our estimated secure bit rates is a lower bound of realistically achievable secure bit rates. Here, we analyze an asymptotic case, where Alice communicates an infinitely long key, $n \rightarrow \infty$. Using this assumption, we can avoid finite-size effects \cite{Leverrier2010} and restrict our analysis to collective Gaussian attacks \cite{Pirandola2008}, while considering them as the most general attacks. 

In collective Gaussian attacks, all physical Gaussian states remain Gaussian states throughout the quantum communication. Additionally, Eve is assumed to interact individually with each state sent by Alice and to store all her extracted states in a quantum memory before applying a joint measurement on her state ensemble at the end of the classical postprocessing.  Then, Eve's eavesdropping attack can be extended, i.e., dilated, into an entangling cloner attack \cite{Grosshans2003_2}. In this attack, Eve starts with a two-mode squeezed (TMS) vacuum state \cite{Fedorov2018} for each incoming state from Alice. One of these modes is coupled to the quantum channel via a beam splitter with transmissivity $\tau_{\mathrm{E}}$, while Eve preserves the other uncoupled mode (for details, see appendix\,\ref{appendix:a}). After interaction with Alice's signal, this coupled mode contains partial information on the communicated key. A ge\-neral TMS state features quantum entanglement, implying that both Eve's modes are strongly correlated. Then, Eve can use these correlations to extract as much information as possible on the sent key.

We model our CV-QKD protocol using an input state $\hat{\rho}_{\mathrm{in}}$ for each symbol $k_i$. This input state contains three modes. The first mode is used by Alice to generate the displaced squeezed states. The remaining two modes describe Eve's TMS state. The first mode of Eve's TMS state locally looks like a thermal state coinciding with the environmental noise state, but possessing quantum entanglement with the third, retained, Eve's mode. Following this formalism, we denote the final state $\hat{\rho}_{\mathrm{out}}$, which is also a three-mode state, where its first mode corresponds to the local state which Bob receives after Eve's attack. Similarly, the remaining two modes describe Eve's modes after her attack. The final state can be written as
\begin{equation}
\label{eq:full_model_general}
\begin{gathered}
\hat{\rho}_{\mathrm{out}} = \hat{T}_{\mathrm{AE}} \, \hat{\rho}_{\mathrm{in}}\,\hat{T}_{\mathrm{AE}}^{\dagger} \text{,}\\
\hat{T}_{\mathrm{AE}}\ = \hat{B}_{\mathrm{AE}}\left( \tau_{\mathrm{E}} \right)\,\hat{D}_{\mathrm{A}}\left(\alpha_{i}\right)\,\hat{S}_{\mathrm{A}}\left(\xi\right) \text{,}
\end{gathered}
\end{equation}
where $\hat{B}_{\mathrm{AE}}\left( \tau_{\mathrm{E}} \right)$ is a beam splitter operator (see appendix\,\ref{appendix:a}) with transmissivity $\tau_{\mathrm{E}}$, which describes coupling Alice's mode to environment (Eve's mode). Similarly, $\hat{D}_{\mathrm{A}}\left(\alpha_{i}\right)$ represents the displacement operator \cite{Fedorov2016} applied to Alice's mode, with the displacement amplitude encoding a specific symbol, $|\alpha_{i}| = |k_{i}|$. Additionally, $\hat{S}_{\mathrm{A}}\left(\xi\right)$ corresponds to the squeeze operator \cite{Pogorzalek2019} acting on Alice's mode, and $\xi$ is the complex squeezing amplitude.

\section{Experimental setup considerations} 
\label{subsec:Experimental_setup}

\begin{figure*}[t]
    \centering
    \includegraphics[width=0.73\textwidth]{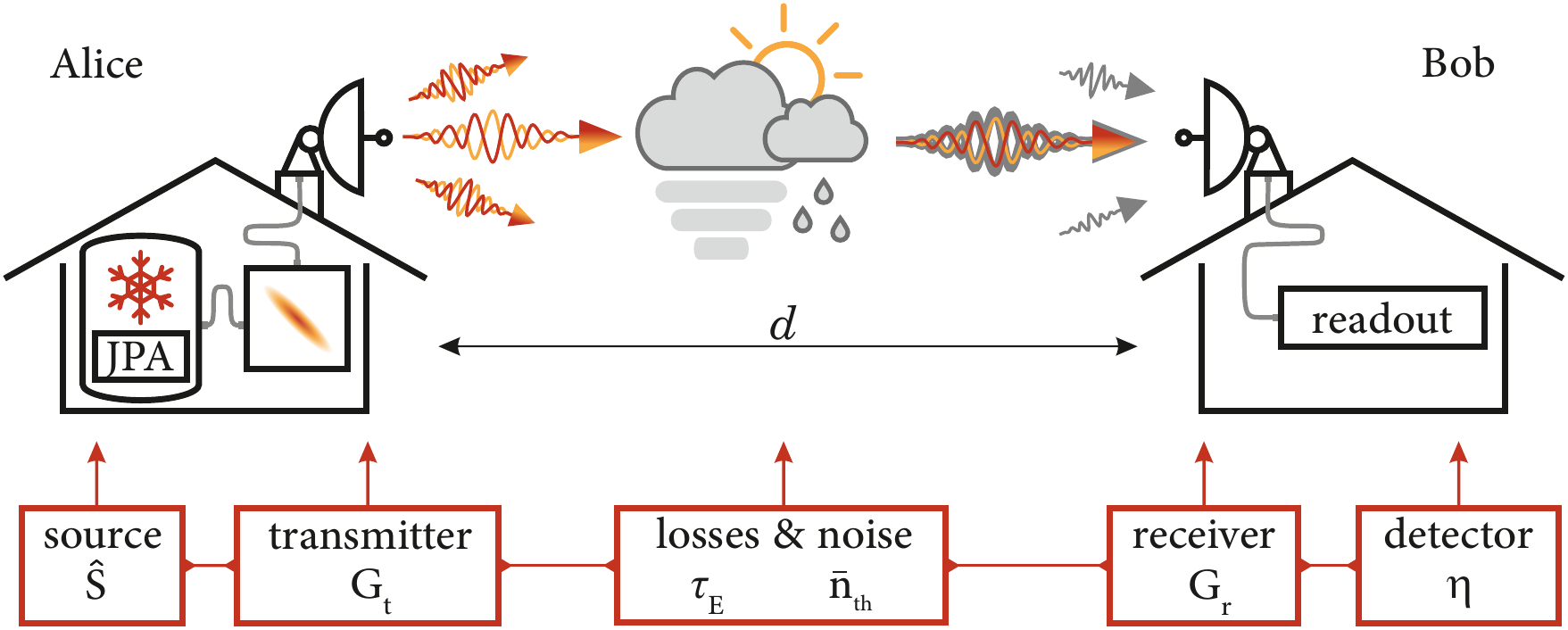}
    \caption{Schematics of main components for an open-air MQC. Source denotes a squeezing generator, typically implemented with a JPA in a cryogenic environment. Transmitter and receiver represent corresponding microwave antennas with gains $G_{\mathrm{t}}$ and $G_{\mathrm{r}}$, respectively. These antennas belong to different communication parties, Alice and Bob, and are separated by a distance $d$. Atmospheric absorption losses are quantified using transmissivity $\tau_{\mathrm{E}}$ which couples the quantum communication channel to the open-air environment with the thermal noise photon number $\bar{n}_{\mathrm{th}}$. Readout is modelled as a homodyne detector with an overall quantum efficiency $\eta$.}
    \label{fig:Fig_2}
\end{figure*}

An open-air implementation of the above introduced CV-QKD protocol requires various different aspects to be taken into account. In this work, we focus on the central components to realize such an open-air MQC and present an associated generic scheme in Fig.\,\ref{fig:Fig_2}. We analyze the generation and detection of propagating squeezed states at millikelvin temperatures. Detection is modelled by a homodyne quadrature measurement with quantum efficiency, $\eta$. Coupling of the squeezed states to the open-air environment (atmosphere) is modelled with two antennas with corresponding gain coefficients. The environment is assumed to be at ambient temperature, $T = \SI{300}{\kelvin}$, and is described by frequency-dependent absorption losses. The latter may also be subject to imperfect weather conditions, as it will be discussed later.
\subsection{Generation of quantum microwave states}  
\noindent

The experimental realization of the CV-QKD protocol requires the generation of propagating displaced squeezed states. In the microwave regime, flux-driven Josephson parametric amplifiers (JPAs) provide an opportunity to generate squeezed states with tunable squeezing level and angle \cite{Yamamoto2008, Zhong2013}. Typically, JPAs consist of a coplanar waveguide $\lambda /4$ resonator which is short-circuited to ground by a direct current superconducting quantum interference device (dc-SQUID). This dc-SQUID acts can be described as nonlinear inductance which can be modulated by applying an external magnetic flux resulting in a flux-tunability of the resonance frequency of the JPA. Applying a microwave pump signal inductively coupled to the $\lambda /4$ resonator, JPAs can provide phase-insensitive or phase-sensitive amplification of incident signals \cite{Yurke1989}. The latter regime is directly related to the generation of squeezed microwave states. According to Cave's theory of noise in li\-near amplifiers \cite{Caves1982}, phase-insensitive bosonic amplifiers are quantum-limited in the sense that they add at least half a noise photon to an input signal. In contrast, a phase-sensitive amplifier can, in principle, achieve noiseless amplification. In practice, JPAs have proven to approach both these limits, which makes them well qualified for MQC applications. In experiments, JPAs operating in the GHz regime were shown to reach noise levels on the order of $0.1$ added noise photons in the phase-sensitive regime \cite{Zhong2013}. Presently, the noise performance of JPAs is limited by fabrication imperfections, pump-induced noise \cite{Renger2021,Fedorov21}, or higher-order nonlinearities \cite{Boutin2017}. The displacement operation required by our CV-QKD protocol can be experimentally realized by applying strong coherent drive tones to cryogenic directional couplers \cite{Fedorov2016}. Ultimately, the combination of JPAs with subsequent directional couplers, allows one to generate displaced squeezed states with a desired displacement amplitude $\alpha$.

\subsection{Microwave antennas} 
\noindent
In order to couple propagating microwave states, generated at millikelvin temperatures, to the open-air quantum channel one requires a microwave interface between the corresponding cryogenic environment and the open-air medium. A microwave antenna serves as such kind of interface. The antenna may be modelled by a transmission line of spatially varying impedance connecting the \SI{50}{\ohm}-matched cryogenic circuits to open-air channels with characteristic impedance of \SI{377}{\ohm}. A central figure of merit of the transmitter and receiver antennas is their passive antenna gain, $G$. In general, for microwave antennas, the gain reads \cite{Pozar2011}
\begin{equation}
    G = \eta_{\mathrm{rad}}\,D \text{,}
\end{equation}
where $0 \leq \eta_{\mathrm{rad}} \leq 1$ is the radiation efficiency and accounts for the antenna losses, while $D$ represents the antenna directivity. The latter expresses the ability of the antenna to focus the emitted power into a specific direction and strongly depends on the antenna geometry. An antenna with a well-defined physical aperture area, $A$, has the directivity
\begin{equation} 
    D = \frac{4\pi A}{\lambda^2}\,e_{\mathrm{A}} \text{,}
\end{equation}
where $A$ is determined by the size and shape of the antenna, $\lambda$ the signal wavelength, and $e_{\mathrm{A}}$ the aperture efficiency, defined as ratio between the effective aperture and physical aperture areas. Cryogenic to open-air transmission of microwave si\-gnals is a current technological challenge. First proposals already exists \cite{gonzalezraya2020coplanar}. For communication distances of around \SI{50}{\meter}, an open-air geometric attenuation of signals, also known as the path loss, is around \SI{80}{\decibel} (see Sec.\,\ref{subsec:Losses_and_noise_budget}) at the frequency of $\SI{5}{GHz}$. In general, the path loss can be compensated by using transmitter and receiver antennas with sufficient gain. For instance, a parabolic transmitter and receiver antennas with a diameter of around $D_{\mathrm{ant}} = \SI{2}{\metre}$ could compensate the aforementioned path loss. A more detailed analysis on antennas designs, gains, and related path losses goes beyond the scope of this paper and is discussed elsewhere \cite{Pirandola2021}. Here, we assume that antenna gains fully compensate for the path loss and focus on the effects of atmos\-pheric absorption losses as the main source of communication imperfections.

\subsection{Quantum efficiency of the detection chain}
\label{subsec:Microwave_detector}

In order to finalize the prepare-and-measure CV-QKD protocol, one has to perform a homodyne quadrature measurement. In the microwave regime, this task requires usage of linear amplifiers with a certain quantum efficiency, $\eta_{\mathrm{mw}}$, to quantify the amplification chain noise performance. The quantum efficiency is defined as the ratio between vacuum fluctuations and fluctuations in output signals resulting from additional noise photons $n_{\mathrm{amp}}$ due to amplification, where $n_{\mathrm{amp}}$ is referred to the input of the detection chain. Therefore, we can express the quantum efficiency as \cite{Renger2021}
\begin{equation}
    \eta_{\mathrm{mw}} = \frac{1}{1+2n_{\mathrm{amp}}} \text{.}
\end{equation}
State-of-the-art travelling wave parametric amplifiers (TWPAs) allow for phase-insensitive amplification with high gain values ($\sim$\,\SI{20}{\deci\bel}) and broad bandwidths ($\sim$\,\SI{3}{\giga\hertz}) at cryogenic temperatures. These TWPAs are also potentially able to approach the quantum-limited regime characterized by $n_\mathrm{amp} = 0.5$ for phase-insensitive mode of operation \cite{Caves1982}. Conversely, as mentioned in Ref.\,\citenum{Castellanos-Beltran2008}, a phase-sensitive linear amplifier could achieve noiseless amplification of a single quadrature, at the cost of deamplifying the conjugate quadrature. Such a detection scheme can be used to implement a microwave homodyne detection similar to its optical counterpart \cite{Eichler2012}. In cryogenic microwave experiments, one typically uses chained quantum-limited amplifiers followed by cryogenic high-electron-mobility transistor (HEMT) amplifiers. In this case, we can use the Friis formula to estimate the total amplification noise $n_{\mathrm{amp}}$ of the detection chain.  For instance, for the case of two chained amplifiers and in the limit large amplification, $G_{1,2} \gg 1$, the total amplification noise reads
\begin{equation}
    n_{\mathrm{amp}} = n_1 + \frac{n_2}{G_1} \text{,}
\end{equation}
where $n_1$ and $G_1$ are the noise photon number and gain of the first amplifier in the chain, while $n_2$ describes the input noise photon number of the second amplifier. Thus, the total noise $n_{\mathrm{amp}}$ depends mainly on the noise properties of the first amplifier. For homodyne detectors at telecom wavelengths, the quantum efficiency is usually modeled by additional losses, introduced by a non-unity transmissivity within a beam splitter model. Both approaches are known to be equivalent as described in Ref.\,\citenum{Boutin2017}. 

\subsection{Losses and noise budget} 
\label{subsec:Losses_and_noise_budget}

We conclude this section with a brief analysis of losses and noise in open-air communication channels, where losses scale with the communication distance. We distinguish between two categories of losses: (i) the path loss which represent geometric attenuation of propagating signals and (ii) absorption losses due to coupling to the environment, such as the atmospheric absorption losses or weather-induced losses. For signals transmitted and received via the antennas, the path loss $L_{\mathrm{p}}$, describing the fraction of the initial signal power lost during the communication, is commonly described using the Friis transmission formula \cite{Pozar2011}
\begin{equation}
\label{eq:Friis_transmission}
    L_{\mathrm{p}} = 10\log_{10} \left( G_{\mathrm{t}} G_{\mathrm{r}} \left (\frac{\lambda}{4\pi d} \right)^{2} \right) \text{.}
\end{equation}
Here, $G_{\mathrm{t}}$ ($G_{\mathrm{r}}$) is the transmitter (receiver) antenna gain, $\lambda$ the wavelength of the communication signals, and $d$ the propagation distance.
The absorption and scattering power losses can be modeled via a single effective beam splitter with transmissivity $\tau_{\mathrm{eff}}$ given by
\begin{equation}
    \tau_{\mathrm{E}} = 10^{-\gamma d/10} \text{,}
\end{equation}
where $\gamma$ is the specific attenuation (dB/km) for a respective loss mechanism. In our case, we attribute these losses to atmospheric absorption and weather conditions, such as rain or haze.
Empirical models show that for microwave frequencies around $f_{\mathrm{mw}} \simeq$\,\SI{5}{\giga \hertz}, propagation losses mainly arise due to molecular oxygen absorption \cite{Ho2004}. For the ideal case of dry weather, we estimate the corresponding specific attenuation of $\gamma_{\mathrm{mw}} = $\,\SI{6.3e-3}{\deci\bel/\kilo\meter} \cite{Ho2004}. To describe the coupling of the propagating quantum bosonic signal $\hat{a}$ to the noisy environmental modes, we use the input-output formalism. The output signal mode $\hat{a}^{\prime}$ can be expressed as
\begin{equation}
\label{environment_coupling}
    \hat{a}^{\prime} = \sqrt{\tau_{\mathrm{E}}}\, \hat{a} + \sqrt{1-\tau_{\mathrm{E}}}\, \hat{h}_{\mathrm{env}} \text{,}
\end{equation}
where $\hat{h}_{\mathrm{env}}$ corresponds to the environmental thermal mode. The latter may be a vacuum or thermal state, depending on the carrier frequency and environment temperature. For a thermal background, the average thermal noise photon number $\bar{n}_{\mathrm{th}}$ per mode is given by the Planck distribution as
\begin{equation}
    \bar{n}_{\mathrm{th}} = \frac{1}{\exp\left(\frac{h\,f}{k_{\mathrm{B}}T} \right) -1} \text{,}
\end{equation}
where $h$ is the Planck constant, $k_{\mathrm{B}}$ is the Boltzmann constant, $f$ the signal frequency, and $T$ the background temperature.

At last, it is instructive to mention open-air losses at telecom wavelengths. We emphasize that Eq.\,\ref{eq:Friis_transmission} is also applicable in the optical frequency range. Then, $G_\mathrm{t}$ and $G_\mathrm{r}$ correspond to the effective passive gain of optical lenses used to focus and collect optical beams. Typical telecom wavelengths (\SI{780}-\SI{850}{\nano\meter}, and \SI{1520}-\SI{1600}{\nano\meter}) are chosen to suffer from the lowest possible atmospheric absorption losses. At the telecom wavelength of \SI{1550}{\nano\meter}, absorption losses of less than \SI{1.0e-2}{\decibel/\kilo\meter}\, can be reached\cite{Kaushal2018}. In this case, open-air attenuation is mainly caused by scattering losses, such as Rayleigh or Mie scattering\, \cite{Kaushal2018}. The corresponding open-air specific attenuation is $\gamma_{\mathrm{tel}} =\SI{2.02e-1}{\deci\bel/\kilo\meter}$. We discuss the additional attenuation due to rain and haze in more detail in Sec.\,\ref{sec:weather_imperfections}.

\section{Security analysis}
\label{subsec:Security}
\subsection{Secret key}
In order to assess the experimental feasibility of our CV-QKD protocol, it is mandatory to analyze its security. The latter is quantified by the secret key $K$, which represents the amount of secure information per communicated symbol and reads as
\begin{equation}
K = \beta \cdot I\left( A \text{:} B\right) - \chi_{\mathrm{E}} \text{.}
\label{SecretKey_DR}
\end{equation}
Here, $I\left( A \text{:} B\right)$ is the mutual information between Alice and Bob and measures correlations between Alice's sent key and Bob's measured key. Additionally, $\chi_{\mathrm{E}}$ is the Holevo quantity \cite{Holevo1973} of Eve and gives an upper bound on the information that Eve obtained during the quantum communication (see appendix\,\ref{appendix:a}). For the sake of simplicity, we assume perfect a reconciliation efficiency $\beta = 1$. It should be noted that experimental values $\beta > 0.9$ have been obtained \cite{Pirandola2018composable}. A positive value of $K$ indicates a secure communication, as Alice and Bob share more information than Eve can in principle obtain.
\begin{figure}[t]
	%\centering
	\includegraphics[width=0.49\textwidth]{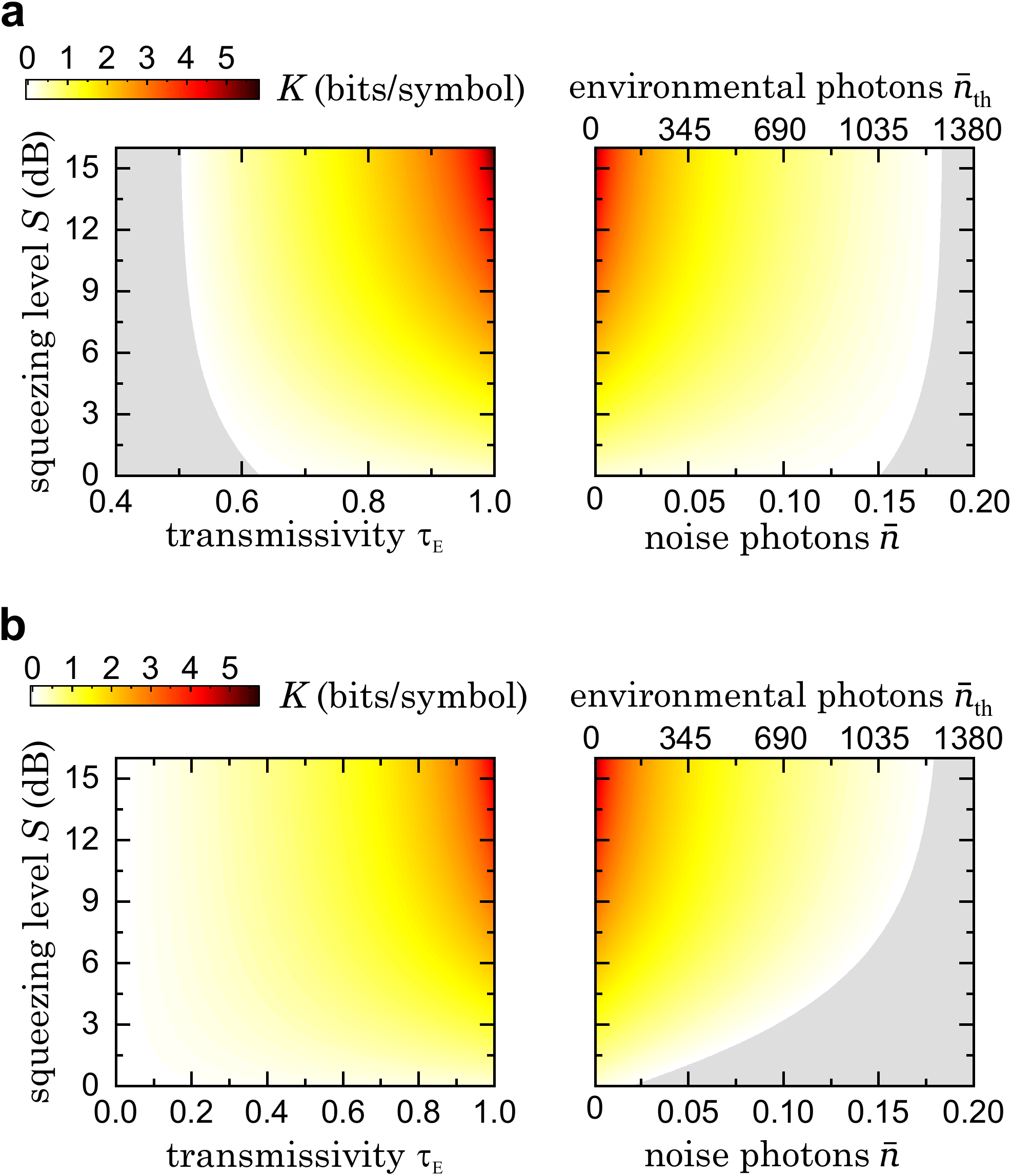}
	\caption{Secret key $K$ of the CV-QKD protocol plotted as a function of the squeezing level $S$ (measured in dB below the vacuum limit), transmissivity $\tau_{\mathrm{E}}$, and noise $\bar{n}$, for ideal detection efficiency $\eta_{\mathrm{mw}} = 1$. Panels \textbf{a} and \textbf{b} show the cases of DR and RR, respectively. Number of environmental noise photons, $\bar{n}$, is shown for a fixed communication distance, $d=\SI{200}{\meter}$, and microwave specific attenuation, $\gamma_{\mathrm{mw}} = $\,\SI{6.3e-3}{\deci\bel/\kilo\meter}. Grey areas represent the regions of negative keys, i.e., insecure communication.}
	\label{fig:Fig_3}
\end{figure}

In Fig.\,\ref{fig:Fig_3} we show results of numerical evaluation of the secret key as a function of the transmissivity $\tau_{\mathrm{E}}$ and noise photons $\bar{n}$ in the quantum channel. Remarkably, in the DR case a secure communication cannot exist when $\tau_{\mathrm{E}}$ exceeds a threshold value of 0.5, which illustrates the well-known result that secure CV-QKD communication in DR schemes are limited to \SI{3}{\deci\bel} of losses \cite{grosshans2002,Weedbrook2012}. The reason for this fact is that communication with DR cannot be secure when Eve receives more than 50\% of Alice's information. In this scenario, Eve effectively replaces Bob as the communication partner. As illustrated in Fig.\,\ref{fig:Fig_3}, this limit can be entirely circumvented by using the RR scheme. Then, Bob's measured key is treated as a reference and Alice needs to correct her own key according to it. For RR, if we imagine that Eve only induces losses during the quantum communication, Alice has always more information than Eve on Bob's measured key. This is because, in her attack, Eve is assumed to induce losses by using a beam splitter to get part of the signals sent by Alice. As a result, Eve can only obtain a fraction of Alice's information. If Eve couples noise photons in addition to the previous losses, the correlations between Alice's sent key and Bob's measured key decrease. At the same time, Eve gains more information on Bob's measured key. In particular, the communication is secure up to a noise photon threshold value $\bar{n}$ of 0.183 for both reconciliation cases. This result is consistent with the well-known Pirandola-Laurenza-Ottaviani-Banchi (PLOB) upper-bounds for Gaussian channels \cite{Pirandola2017}. This noise threshold corresponds to the crossover of the quantum channel capacity from finite values to zero. It is also important to note that these noise numbers do not account for noise photons which can be added by Bob during his measurements. Finally, we observe that an increase in squeezing level results in an increase of the secret key. This increase can be understood as a decrease of the displacement uncertainty encoding the symbols, while also allowing for higher displacement amplitudes \cite{Cerf2001}.
\begin{figure}[t]
	%\centering
	\includegraphics[width=0.48\textwidth]{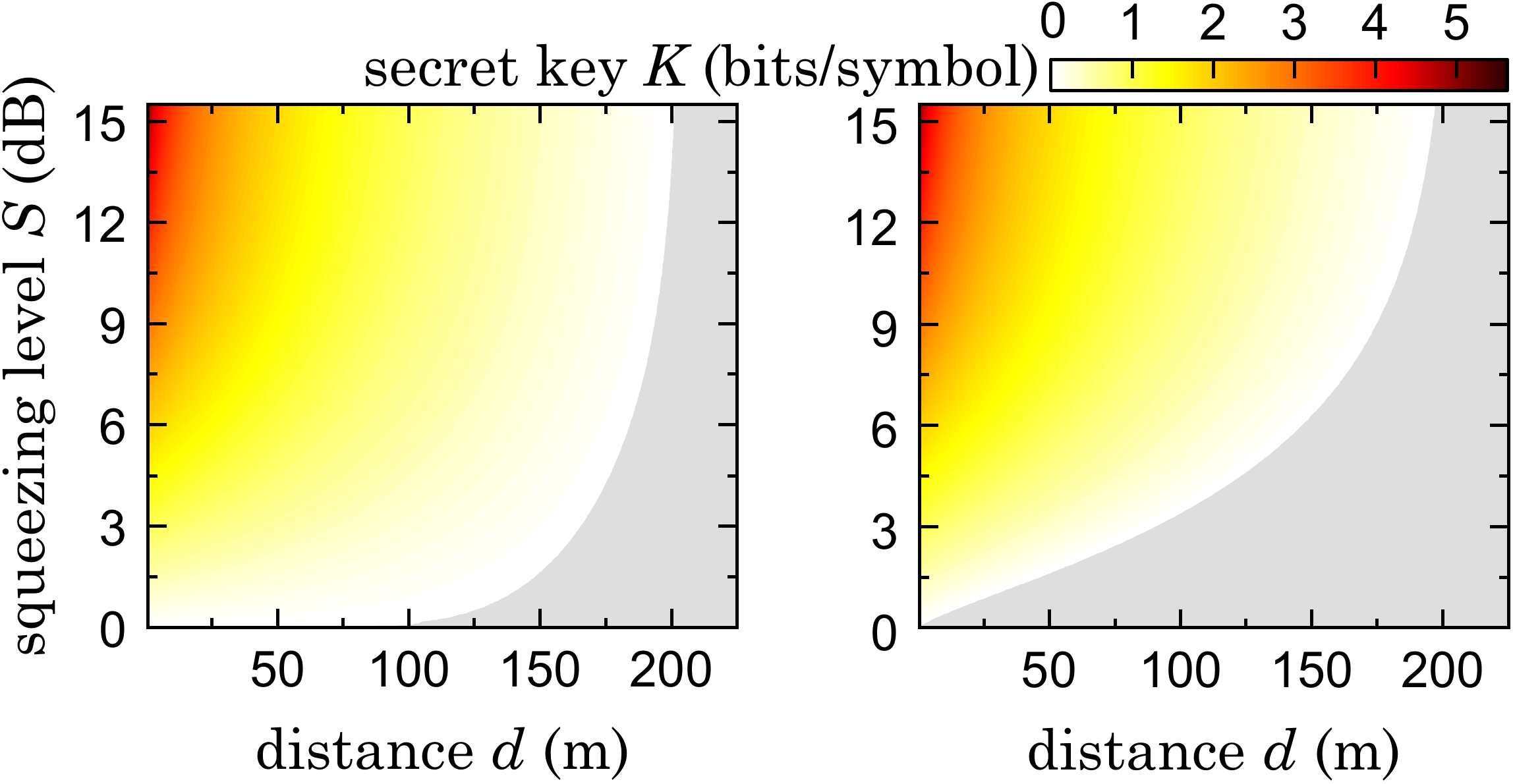}
	\caption{Secret key $K$ of the CV-QKD protocol as a function of communication distance $d$ and squeezing level $S$. Left (right) plot corresponds to the DR (RR) cases, respectively. Squeezing is given in dB below the vacuum level. Detection efficiency is assumed to be ideal, $\eta_\mathrm{mw} = 1$. We assume the average environmental noise photon number $\bar{n}_{\mathrm{th}}  = 1250$ and transmission losses $\gamma_{\mathrm{E}} = \gamma_{\mathrm{mw}} \simeq \SI{6.3e-3}{\deci\bel/\kilo\meter}$. Grey areas represent the regions of negative keys, i.e., insecure communication.} 
	\label{fig:Fig_4}
\end{figure}

Next, we extend our analysis of the microwave CV-QKD protocol to variable communication distances $d$, for both DR and RR. To this end, we use Eq.\,\ref{environment_coupling} in combination with the specific attenuation given in Sec.\ref{subsec:Losses_and_noise_budget} to convert communication distances $d$ into corresponding transmissivities $\tau_{\mathrm{E}}$. The corresponding secret keys are shown in Fig.\,\ref{fig:Fig_4}. Remar\-kably, we observe positive secret key values over communication distances of up to \SI{200}{\meter}, in both DR and RR. These results suggest the experimental feasibility of microwave QKD in open-air conditions. No major distinction in communication distances is observed between the reconciliation cases, although one could intuitively expect RR to yield larger distances according to our previous discussion. This behavior originates from the presence of the bright microwave thermal background which couples to propagating states during the communication. Consequently, the effects of coupled noise largely outweigh the effects of losses and make the RR and DR cases more similar.

For a more practical evaluation of the QKD performance, one typically uses a secret key rate $R_{0}$. The latter evaluates the amount of secure bits per second that can be obtained from the communication protocol. Under the asymptotic case assumption, one can express the secret key rate $R_{0}$ in bits per second using the secret key as
\begin{equation}
\label{equation:secret_key_rate}
    R_{0} = f_{\mathrm{r}} \cdot K \text{,}
\end{equation}
where $f_{\mathrm{r}}$ represents the repetition rate (in symbols per second). This rate which encompasses all information post-processing steps, such as sifting, parameter estimations \cite{Scarani2009,Laudenbach2018}, and experimental bandwidths of the involved devices. We use an upper bound on the secret key rate, $R$, derived from the Shannon-Hartley theorem and the Nyquist rate \cite{nyquist1928}
\begin{equation}
\label{equation:secret_key_rate_upper_bound}
    R_{0} \leq R = 2\,\Delta f \cdot K \text{,}
\end{equation}
where $\Delta f$ denotes the experimental detection bandwidth. This upper bound becomes especially useful when comparing different physical QKD platforms, as it is going to be discussed in the next section. \\

\begin{figure}[t]
	%\centering
	\includegraphics[width=0.49\textwidth]{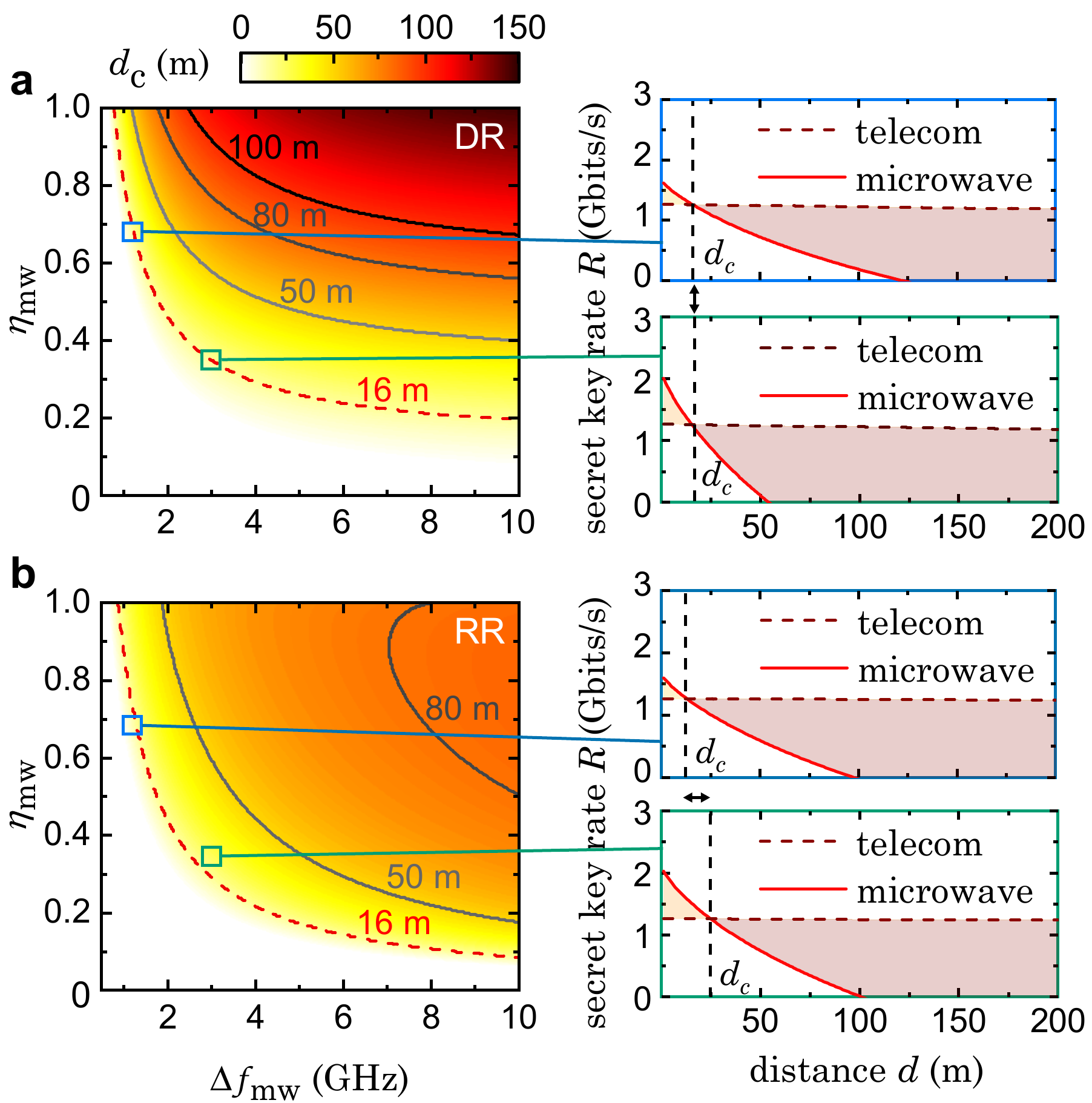}
	\caption{Crossover distance $d_{\mathrm{c}}$ between microwave and telecom CV-QKD. Panels \textbf{a} and \textbf{b} illustrate the DR and RR cases, respectively. For the telecom and microwave wavelengths, we assume transmission losses $\gamma_{\mathrm{tel}} \simeq \SI{2.02e-1}{\deci\bel/\kilo\meter}$, and $\gamma_{\mathrm{mw}} \simeq \SI{6.3e-3}{\deci\bel/\kilo\meter}$, respectively. For both DR and RR, the secret key rates $R$ of both detection cases are shown on the right co\-lumn as a function of communication distance $d$.}
	\label{fig:Fig_5}
\end{figure}

\subsection{Comparison between telecom and microwave frequency} 

Here, we now compare the microwave CV-QKD performance to that of QKD at telecom frequencies. For this purpose, we define and numerically compute a communication crossover distance $d_{\mathrm{c}}$
\begin{equation}
    d_{\mathrm{c}} \coloneqq \max_{R_{\mathrm{mw}} \geq R_{\mathrm{tel}}} \left(d \right) \text{,}
\end{equation}
where $d$ corresponds to communication distance, while $R_{\mathrm{mw}}$ and $R_{\mathrm{tel}}$ are the secret key rates for microwave and telecom frequencies, respectively. According to Eq.\,\ref{equation:secret_key_rate_upper_bound} , it is relevant to optimize the detection bandwidth to achieve high secret key rates. To this end, we assume an experimental state-of-the-art broadband squeezing generation and detection at \SI{1550}{\nano\meter} wavelength over a bandwidth of $\Delta f_{\mathrm{tel}} = \SI{1.2}{\giga\hertz}$ with a quantum efficiency of $\eta_{\mathrm{tel}} = 0.53$, as shown in Ref.\,\citenum{Ast2013}. In this experiment, the authors also report a squeezing level of \SI{3}{\deci\bel}, which we will use as a common level of vacuum squeezing for both the microwave and telecom regimes. We compute the corresponding crossover distance as a function of the microwave detection bandwidth $\Delta f_{\mathrm{mw}}$ and quantum efficiency $\eta_{\mathrm{mw}}$. The corresponding results are shown in Fig.\,\ref{fig:Fig_5} for both DR and RR. Interestingly, we observe that the microwave CV-QKD protocol can outperform the telecom counterpart for realistic values of $ \Delta f_{\mathrm{mw}}$ and $\eta_{\mathrm{mw}}$. A clear distinction can be seen between the two reconciliation cases. For the DR case, it is beneficial to aim at the quantum efficiency close to unity and large detection bandwidths. The situation is noticeably different in RR. For the latter, we observe that above a certain detection bandwidth, the optimal quantum efficiency is no longer unity. Instead, there exists an optimal detection noise added by Bob, which maximizes the secret key rate depending on the detection bandwidth. The existence of an optimal quantum efficiency is a remarkable feature of RR, which arises when Bob couples additional noise during his measurements \cite{Garcia-Patron2009}. To illustrate the influence of the quantum efficiency and the detection bandwidth, we envision two different microwave homodyne detection cases implemented by a phase-sensitive amplifier. First, we choose a high detection bandwidth $\Delta f_{\mathrm{mw}} = \SI{3}{GHz}$  with a quantum efficiency of $\eta_{\mathrm{mw}} = 0.345$. This case is motivated by the existing state-of-the-art superconducting TWPA devices operated in the phase-insensitive regime \cite{Macklin2015,Perelshtein2021}. The second case considers a detection bandwidth of $\Delta f_{\mathrm{mw}} = \SI{1.2}{GHz} = \Delta f_{\mathrm{tel}}$, and we choose a quantum efficiency of $\eta_{\mathrm{mw}} = 0.695$, such that both cases yield the same DR crossover distance. This case originates from recent results on broadband squeezing in the microwave regime \cite{Schneider2020,Qiu2022}. By using this set of already experimentally feasible parameters, we can reach a crossover distance of $d_{\mathrm{c}} = \SI{16}{\meter}$ for both cases. For RR, we observe that the crossover distance can be increased to $d_{\mathrm{c}} = \SI{25}{\meter}$. The reason is that RR benefits from a quantum efficiency below unity. Remarkably, high secret key rates $R$ of a few Gbits per second can be reached for all of the previously mentioned set of parameters. However, we stress that the computed secret key rates $R$ are merely upper bounds for realistically achievable rates. Existing telecom QKD implementations reach secure key rates up to a few Mbits per second \cite{Huang2015,Tang2020,Sarmiento2022}. Aside from finite quantum detection efficiencies and bandwidths, practical secret key rates are also limited by various factors such as actual experimental repetition rates \cite{Laudenbach2018}, device-induced noise \cite{Tang2020}, finite size effects \cite{Leverrier2010}, or post-processing \cite{Lodewyck2007}. Nevertheless, the demonstrated results make MQC relevant for short-distance classical communication protocols such as Wifi 802.11 standard (maximal communication distance $\simeq \SI{70}{\m}$), Bluetooth 5.0 ( $\simeq \SI{240}{\m}$), or more recent technologies such as 5G ($\simeq \SI{305}{\m}$) because of matching frequency ranges, distances, and technological infrastructure. 

\section{Weather induced losses effects}
\label{sec:weather_imperfections}

\subsection{Non-optimal weather conditions}
\label{subsec:weather_conditions}
So far, we have investigated open-air CV-QKD under ideal weather conditions. However, it is well-known that realistic, non-optimal weather conditions may drastically affect absorption losses for propagating signals. Such effects are especially prominent in the telecom frequency range. Therefore, it is na\-tural to investigate effects of these non-optimal conditions on the MQC as well. Specifically, we focus on two non-ideal weather scenarios: rain and haze.
\begin{figure*}[t]
	%\centering
	\includegraphics[width=0.85\textwidth]{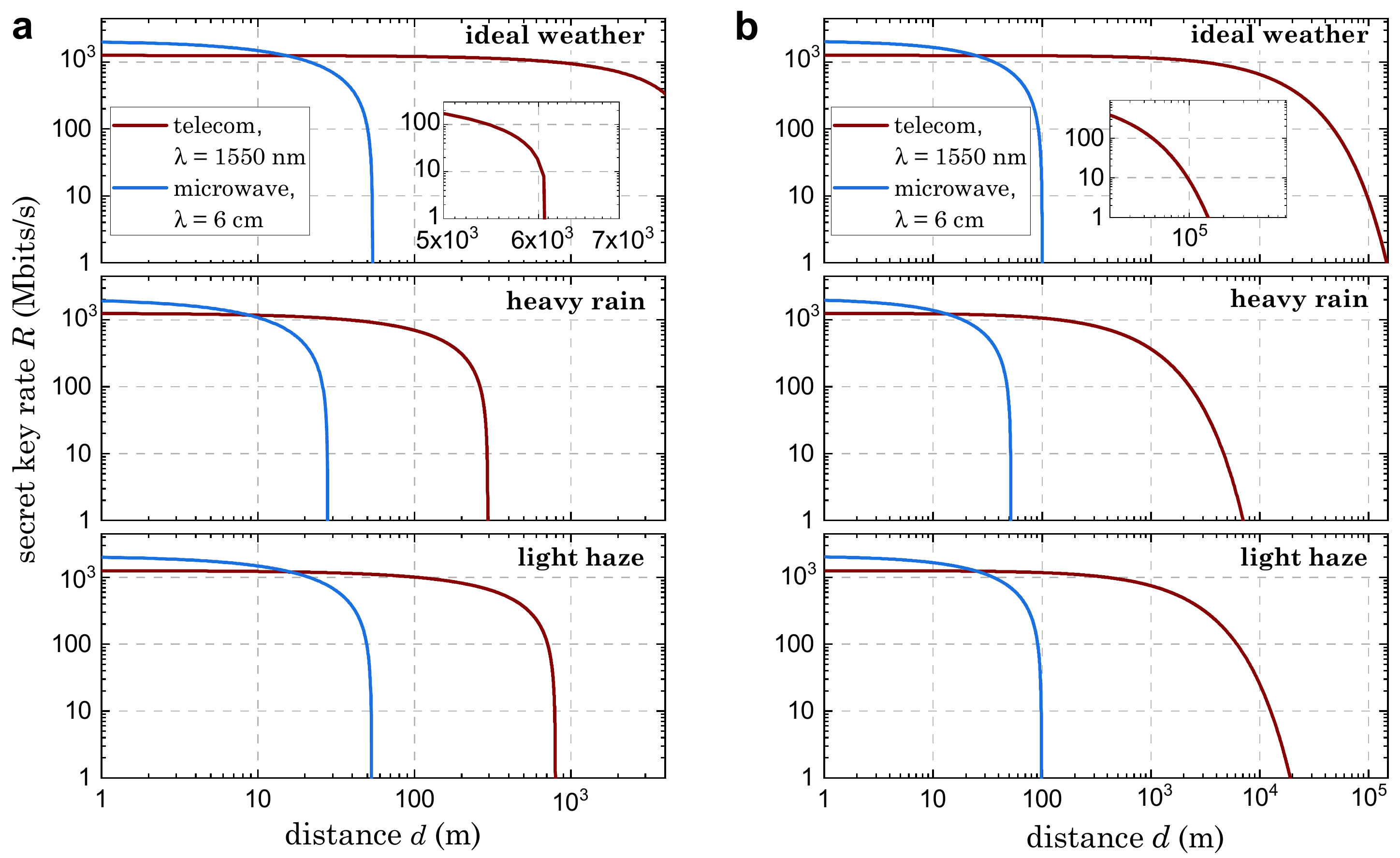}
	\caption{Secret key rates of the CV-QKD protocol for various weather conditions. Telecom, and microwave secret key rates $R$ are computed in DR (panel $\textbf{a}$) and in RR (panel $\textbf{b}$) as a function of the communication distance $d$ for the squeezing levels of $S_{\mathrm{tel}} = S_{\mathrm{mw}} = \SI{3}{\deci\bel}$. Three different weather conditions are considered: ideal weather conditions (visibility of \SI{23}{\kilo\meter}), heavy rain with a rain rate of \SI{7}{\milli\meter/h}, and light haze with a visibility of \SI{4}{\kilo\meter}. The choice of quantum efficiency and detection bandwidth is the same as for the ideal weather conditions. For the telecom case, we consider the total transmission losses $\gamma_{\mathrm{tel}} \simeq \SI{2.02e-1}{\decibel/\kilo\meter}$ (optimal), $\gamma_{\mathrm{tel}} \simeq \SI{4.17}{\decibel/\kilo\meter}$ (rain), and $\gamma_{\mathrm{tel}} \simeq \SI{1.55}{\deci\bel/\kilo\meter}$ (haze). For the microwave case, we assume the transmission losses $\gamma_{\mathrm{mw}} \simeq \SI{6.3e-3}{\decibel/\kilo\meter}$ (optimal), $\gamma_{\mathrm{mw}} \simeq \SI{1.22e-2}{\deci\bel/\kilo\meter}$ (rain), and $\gamma_{\mathrm{mw}} \simeq \SI{6.4e-3}{\deci\bel/\kilo\meter}$ (haze).}
	\label{fig:Fig_6}
\end{figure*}
In the context of microwave communication, the ITU-P.\,838 and ITU-P.\,840 recommendations provide empirical prediction models for the induced attenuation on propagating microwave signals due to rainfall and haze, respectively. More precisely, the specific attenuation $\gamma_{\mathrm{mw,r}}$ due to rain along a horizontal path can be expressed as \cite{Shrestha2019}
\begin{equation}
\label{losses_rain_mw}
\gamma_{\mathrm{mw,r}} = k\left(f\right)\, R_{\mathrm{r}}^{\alpha\left(f\right)} \text{,} 
\end{equation}
where $k$ and $\alpha$ are coefficients which depend on the communication microwave frequency $f$, while $R_{\mathrm{r}}$ (mm/h) is the rain rate. The haze specific attenuation $\gamma_{\mathrm{h}}$ can be obtained from the li\-quid water concentration $M$ (g/$\mathrm{cm}^{3}$) from a linear relationship as \cite{Zhao2000}
\begin{equation}
\label{losses_haze_mw}
\gamma_{\mathrm{mw,h}} = K_{\mathrm{l}}\left(f,T\right)\,M \text{,}
\end{equation}
where $K_{\mathrm{l}}$ ((dB/km) / (g/$\mathrm{cm}^{3}$)) is the linear attenuation that depends on the considered microwave frequency $f$ and water temperature $T$ in the atmosphere. The liquid water concentration can be related to a physically more intuitive quantity, the so-called visibility $V$ (km). The latter represents the distance at which the light intensity from an object drops to 2$\%$ of its initial value. For a non-polluted environment, one can link two aforementioned quantities as \cite{Fisak2006} 
\begin{equation}
\label{liquid_water_concentration}
M = \left(\frac{a}{V}\right)^{b} \text{,}
\end{equation}
where $a = -\log\left(0.02\right)/99$ and $b = 0.92^{-1}$. For the telecom frequencies, rain causes a wavelength-independent attenuation. The specific attenuation $\gamma_{\mathrm{tel,r}}$ can be expressed for a horizontal path as \cite{Kaushal2018}
\begin{equation}
\label{losses_rain_MIR}
\gamma_{\mathrm{tel,r}} = k\, R_{\mathrm{r}}^{\alpha} \text{,} 
\end{equation}
where $R_{\mathrm{r}}$ is the rain rate, $k = 1.076$, and $\alpha = 0.67$. The haze-specific attenuation is empirically derived similarly to the microwave case. Once again, visibility determines the specific attenuation $\gamma_{\mathrm{tel,r}}$. Empirical models for Mie scattering show that \cite{Kim2001,Kaushal2018}
\begin{equation}
\label{losses_haze_tel}
\gamma_{\mathrm{\lambda,h}} =  \frac{C}{V}\, \left(\frac{\lambda}{550} \right)^{-p\left(V\right)} \text{,}
\end{equation}
where $C = 39.1\,\log\left(e\right)$, $\lambda$ (nm) corresponds to a certain telecom wavelength, and $p$ is a scattering coefficient that depends on the considered visibility range and varies from 0 to 1.6. \cite{Kim2001,Kaushal2018}.\\

\subsection{Effects of weather conditions} 

In order to study the effect of non-optimal weather conditions on the CV-QKD secure key rates, we consider two specific situations: $\left(\text{i}\right)$ heavy rain with the rate $R_{\mathrm{r}} = 7$ mm/h and $\left(\text{ii}\right)$ light haze with a visibility $V = \SI{4}{\kilo\meter}$. We compare the telecom and microwave secret key rates in Fig.\,\ref{fig:Fig_6}. For the detection bandwidth and quantum efficiency, we stick to the previously analyzed set of parameters ($\Delta f_{\mathrm{tel}} =$ \SI{1.2}{GHz}, $\eta_{\mathrm{tel}} =0.53$ and $\Delta f_{\mathrm{mw}} =$ \SI{3}{GHz}, $\eta_{\mathrm{mw}} =0.345$). We find that short-distance microwave QKD could potentially yield higher secret key rates as compared to the telecom case. The reason is that microwave QKD benefits from higher experimental bandwidths and lower losses due to weather imperfections. We note that telecom QKD allows for secure communication over much larger distances, up to $d \simeq \SI{140}{\kilo\meter}$ using RR. These distances are significantly reduced when the effects of rain and haze are considered. For these weather conditions, the maximum secure telecom communication distances are strongly reduced to \SI{300}{\meter} (\SI{7}{\kilo\meter}) and \SI{800}{\meter} (\SI{2}{\kilo\meter}), in the DR $\left(\text{RR}\right)$ respectively. Conversely, for microwave frequencies, the maximum secure communication distance is almost unchanged in both reconciliation cases compared to that obtained for optimal weather conditions, highlighting the robustness of microwave CV-QKD to weather effects. The most significant difference arises when considering the effect of light haze. Remarkably, haze induces little-to-no microwave losses. Even strong haze and fog only weakly disturbs microwave signals by creating a small additional attenuation of around $10^{-3}$ dB/km. The latter holds even when visibility is reduced to less than \SI{500}{\m}. In contrast, reducing the visibility below $\SI{4}{\kilo\meter}$ would generate large losses (more than $\SI{10}{\deci\bel/\kilo\meter}$) for the telecom frequency, preventing the possibility of any relevant secure quantum communication. These results indicate that an ideal quantum open-air communication network could consist of a combination of microwave-based channels for short distances ($d \leq \SI{200}{\meter}$) and telecom-based channels for long distances ($d > \SI{200}{\meter}$).

\section{Discussion}
In conclusion, we have performed a comprehensive analysis of microwave CV-QKD and demonstrate its potential for applications in open-air conditions. We have shown that quantum microwaves can yield positive secret key rates for short-distance communication for both DR and RR. Our calculations rely on empirical models for microwave and telecom atmospheric absorption losses. We have estimated the related microwave and telecom specific attenuation for optimal weather conditions to \SI{6.3e-3}{\decibel / \kilo\meter} and \SI{2.02e-1}{\decibel / \kilo\meter}, respectively. In our analysis, we have assumed microwave homodyne detection based on state-of-the-art TWPAs. Our model for the CV-QKD protocol predicts positive secret key rates for the microwave regime over distances of around \SI{200}{\meter}. We have employed this model to compare the microwave and telecom cases for different detection quantum efficiencies and bandwidths. Our results show that, based on parameters of state-of-the-art technology, the microwave CV-QKD can potentially outperform the telecom implementations for short distances of around $\SI{30}{\meter}$ in terms of the secret key rates. From our analysis, it appears that both reconciliation scenarios are relevant. In particular, DR is favored for high quantum efficiencies, while RR allows for applications with rather lower detection quantum efficiencies $\eta$. The RR case also exhibits a nontrivial dependence of the secret key rate R on $\eta$, which can be explained by the positive impact of detection noise on the protocol security.

Finally, we have considered the open-air CV-QKD protocol under non-ideal weather conditions of rain and haze. We have found that these non-idealities strongly reduce the secure communication distance for the telecom regime, from $\SI{140}{\kilo\meter}$ to several hundred meters. Remarkably, the microwave open-air CV-QKD protocol appears to be largely immune to these weather imperfections with its secure communication distances staying mostly unchanged. We envision that our results to serve as a motivation for building first prototypes of secure microwave quantum local area networks. Our analysis also establishes the foundations for hybrid networks, where short-distance secure communication is carried out by microwave signals. Such hybrid network offer the advantage of providing potential high secret key rates and robustness to weather imperfections, while switching to telecom setups for long-distance communication. Short-distance MQC secure platforms could also complement current classical microwave communication technologies such as Wifi, Bluetooth, and 5G due to the intrinsic frequency and range compatibilities.

\appendix
\section{Description of Bob's and Eve's quantum states}
\label{appendix:a}
In this section, we provide details about the states of Bob and Eve. To describe Bob's and Eve's quantum states, we first need to consider the matrix representing the beam splitter operator

\begin{equation}
\mathcal{B}\left(\tau \right) = 
\begin{pmatrix}
\sqrt{\tau}\,\textbf{I}_{\mathrm{2}}\,, & \sqrt{1-\tau}\,\textbf{I}_{\mathrm{2}} \\
-\sqrt{1-\tau}\,\textbf{I}_{\mathrm{2}}\,, & \sqrt{\tau}\,\textbf{I}_{\mathrm{2}}\\
\end{pmatrix} \text{.}
\end{equation}
Here, $\tau$ is the transmissivity associated with the beam splitter and $\textbf{I}_{\mathrm{2}}$ denotes the 2$\times$2 identity matrix.
Further, we introduce the direct sum for matrices $\textbf{A}$ and $\textbf{B}$ as 
\begin{equation}
\textbf{A} \oplus \textbf{B} = 
\begin{pmatrix}
\textbf{A}\,, & \textbf{0} \\
\textbf{0}\,, & \textbf{B}
\end{pmatrix} \text{.}
\end{equation}
Since we assume all states in the protocol to be Gaussian, these states are characterized by their displacement vector $\hat{x}$ and covariance matrix $\textbf{V}$ \cite{Laudenbach2018}. In this formalism, the displacement vector of an $N$-mode Gaussian state reads as
\begin{equation}
\hat{x} = \left(\hat{q}_{\mathrm{1}},\hat{p}_{\mathrm{1}},\dots,\hat{q}_{\mathrm{N}},\hat{p}_{\mathrm{N}} \right)  \text{,}
\end{equation}
where $\hat{q}_{i},\hat{p}_{i}$ are the conjugate quadrature operators of the $i_{\mathrm{th}}$ mode. The displacement vector fulfills the commutation relation
\begin{equation}
\begin{gathered}
\left[\hat{x}_{i},\hat{x}_{j}\right] = \frac{i}{2}\boldsymbol\Omega_{ij} \text{,} \\
\boldsymbol\Omega = \bigoplus_{i\mathrm{=1}}^{N}
\begin{pmatrix}
0\,, & 1 \\
-1\,, & 0
\end{pmatrix} \text{.}
\end{gathered}
\end{equation}
We use the expression $\bar{\textbf{x}}$ to refer to the expectation value of the displacement vector, i.e.,
\begin{equation}
    \bar{\textbf{x}} = \langle \hat{x} \rangle \text{.}
\end{equation}
The elements of the covariance matrix of a mode are computed as
\begin{equation}
\textbf{V}_{ij} = \langle \hat{x}_{i}\hat{x}_{j} + \hat{x}_{i}\hat{x}_{j} \rangle /2 - \langle\hat{x}_{i}\rangle\,\langle\hat{x}_{j} \rangle \text{.}
\end{equation}
Using the previously introduced matrices in combination with Eq.\,\ref{eq:full_model_general}, we can express the mean displacement vector of Bob's mode (Eve's mode) $\bar{\textbf{x}}_{\mathrm{B}}$ ($\bar{\textbf{x}}_{\mathrm{E}}$) as well as the covariance matrix of Bob's mode (Eve's mode) $\textbf{V}_{\mathrm{B}}$ ($\textbf{V}_{\mathrm{E}}$) as
\begin{equation}
\label{eq:Bob_and_Eve_states}
\begin{gathered}
\left(
\bar{\textbf{x}}_{\mathrm{B}} \text{,}\thickspace
\bar{\textbf{x}}_{\mathrm{E}} 
\right)^\mathrm{T} =
\mathbf{\Sigma} \cdot
\left(
\bar{\textbf{x}}_{\mathrm{0}} \text{,}\thickspace
\bar{\textbf{x}}_{\mathrm{E,in}} 
\right)^\mathrm{T} + \mathbf{\Sigma_{\mathrm{E}}} \cdot \left(
\bar{\textbf{x}}_{\mathrm{A}} \text{,}\thickspace
\bar{\textbf{0}}_{\mathrm{E,in}} 
\right)^\mathrm{T} 
\text{,}\\
\begin{pmatrix}
\textbf{V}_{\mathrm{B}}\,, & \textbf{C}_{\mathrm{BE}} \\
\textbf{C}_{\mathrm{BE}}^\mathrm{T}\,, & \textbf{V}_{\mathrm{E}}
\end{pmatrix}
=
\mathbf{\Sigma} \cdot \left(\textbf{V}_{\mathrm{0}} \oplus \textbf{V}_{\mathrm{E,in}} \right) \cdot \mathbf{\Sigma}^\mathrm{T} \text{,}
\end{gathered}
\end{equation}
with
\begin{equation}
\begin{gathered}
\mathbf{\Sigma_{\mathrm{E}}} = \mathcal{B}\left(\tau_{\mathrm{E}}\right) \oplus \textbf{I}_{\mathrm{2}} \text{,} \\
\mathbf{\Sigma_{\mathrm{A}}} = \textbf{R}\left(\varphi/2\right) \cdot \textbf{S}\left(r\right) \oplus \textbf{I}_{\mathrm{4}} \text{,} \\
\mathbf{\Sigma} = \mathbf{\Sigma_{\mathrm{E}}} \cdot \mathbf{\Sigma_{\mathrm{A}}} \text{.}
\end{gathered}
\end{equation}
Here, the dot $\cdot$ represents a matrix multiplication and $\bar{\textbf{x}}_{\mathrm{0}}$ ($\textbf{V}_{\mathrm{0}}$) the mean displacement vector (covariance matrix) of the initial vacuum state. Furthermore, $\bar{\textbf{x}}_{\mathrm{A}}$ represents Alice's mean displacement vector, and $\bar{\textbf{x}}_{\mathrm{E,in}}$ represents Eve's initial mean displacement vector of her TMS state. Additionally, $r$ and $\varphi$ correspond to the squeezing factor and squeezing angle of the generated squeezed states by Alice, respectively. Correlations between Bob's and Eve's individual states are described by the submatrix $\textbf{C}_{\mathrm{BE}}$. Finally, $\textbf{R}$ corresponds to a 2D rotation matrix while $\textbf{S}_{\mathrm{sq}}$ is a 2$\times$2 matrix, which we calculate as
\begin{equation}
\begin{aligned}
&\textbf{R}\left(\varphi/2\right)\cdot\textbf{S}_{\mathrm{sq}}\left(r\right) = \\
    &\begin{pmatrix}
\mathrm{cos}\left(\varphi/2\right)\,, & \mathrm{sin}\left(\varphi/2\right) \\
-\mathrm{sin}\left(\varphi/2\right)\,, & \mathrm{cos}\left(\varphi/2\right)\\
\end{pmatrix} \cdot
\begin{pmatrix}
\mathrm{exp}\left(-r\right)\,, & 0 \\
0\,, & \mathrm{exp}\left(r\right)\\
\end{pmatrix} \text{.}
\end{aligned}
\end{equation}
Using Eq.\,\ref{eq:Bob_and_Eve_states}, the variance of Bob's states reads
\begin{equation}
\begin{aligned}
    \textbf{V}_{\mathrm{B}} &= \tau_{\mathrm{E}} \textbf{V}_{\mathrm{A}} + \left( 1 - \tau_{\mathrm{E}} \right) \frac{1}{4} \left( 1 + 2 n_{\mathrm{Eve}} \right) \textbf{I}_{\mathrm{2}} \\
    & = \tau_{\mathrm{E}} \textbf{V}_{A} + \left[\frac{1}{4} \left( 1 - \tau_{\mathrm{E}} \right) + \bar{n} \right] \textbf{I}_{\mathrm{2}} \text{,}
\end{aligned}
\end{equation}
where $\textbf{V}_{\mathrm{A}}$ represents Alice's state variance while the covariance matrix of Eve's mode coupled to Alice's mode is given by $0.25 \left( 1 + 2 n_{\mathrm{Eve}} \right)\textbf{I}_{\mathrm{2}}$. Lastly, we incorporate the noise of the amplification chain by using the following input-output forma\-lism for a bosonic signal mode $\hat{a}$ \cite{Caves1982}
\begin{equation}
    \hat{a}^{'} = \sqrt{G}\,\hat{a} + \sqrt{G-1}\,\hat{h}_{\mathrm{amp}}^{\dagger} \text{.}
\end{equation}
Here, $G$ is the gain of the amplification chain and $\hat{h}_{\mathrm{amp}}$ is an environmental mode, modelled as a thermal state. For $G \gg 1$, this results in the final covariance matrix for Bob:
\begin{equation}
    \textbf{V}_{\mathrm{B}} = \tau_{\mathrm{E}} \textbf{V}_{A} + \left[\frac{1}{4} \left( 1 - \tau_{\mathrm{E}} \right) + \bar{n} + \bar{n}_{\mathrm{g}}\right] \textbf{I}_{\mathrm{2}} \text{.}
\end{equation}
Here, $\bar{n}_{\mathrm{g}} = \bar{n}_{\mathrm{amp}} /2$ is the added quadrature noise from the amplification chain expressed in an average photon number.   
In the previous equation, the covariance matrix has been divided by the gain $G$ as this gain can always be determined from calibration measurements.
In the third step, we compute the mutual information $I\left( A \text{:} B\right)$ using the expression
\begin{equation}
\label{mutual_information}
I\left( A \text{:} B\right) = h\left(B\right) - h\left(B\vert A \right) \text{,}
\end{equation}
where $h$ denotes the differential entropy. Local measurements of Bob on individual states he receives during the communication are represented by a classical random variable $B \vert A$. Then, $B$ is a classical random variable representing Bob's overall measurements over all received states (i.e., represen\-ting Bob's final key estimation $\mathcal{K}' = \left\lbrace k_{\mathrm{1}}', \dots, k_{\mathrm{N}}' \right\rbrace$ ).
Furthermore, to compute Holevo quantity in DR $\chi_{\mathrm{E,DR}}$ and in RR $\chi_{\mathrm{E,RR}}$, we first start by finding Eve's average state. To this end, we introduce an integer $c \in \{ 0\, , 1 \}$ describing the choice of Bob's measurement basis, $q$ or $p$. Note that due to the sifting step, Bob's measurement basis matches Alice's encoding basis. Eve's average state reads as
\begin{equation}
\hat{\rho}_{\mathrm{avg,E}} = \sum_{\mathrm{c = 0,1}} \int_{-\infty}^{\infty} f_{\mathrm{A,C}}\left(k_{i},c\right) \, \hat{\rho}_{\mathrm{E}}^{k_{i}} \, \mathrm{d}k_{i} \text{,}
\end{equation}
where $f_{\mathrm{A,C}}\left(k_{i},c\right)$ is the probability of Alice encoding a symbol $k_{i}$ in a measurement basis according to $c$. Additionally, $C$ represents a binary random variable used to obtain $c$ with the same probability for both outcomes ($P\left(C = 0\right) = P\left(C = 1\right) = 1/2$). Since $c$ is given by a discrete variable and $k_{i}$ by a continuous variable, we use a mixed joint probability density function which gives
\begin{equation}
\begin{aligned}
f_{\mathrm{A,C}}\left(k_{i},c\right) &= f_{\mathrm{A\vert C}}\left(k_{i}\vert c\right)p\left(C = c \right) \\
&= f_{\mathrm{A}}\left(k_{i}\right)p\left(C = c \right) \\
&= \frac{1}{\sqrt{2 \pi \sigma_{\mathrm{A}}^2}}\exp\left( -\frac{k_{i}^2}{2 \sigma_{\mathrm{A}}^2}\right) \frac{1}{2} \text{,}
\end{aligned}
\end{equation}
where $\hat{\rho}_{\mathrm{E}}^{k_{i}}$ is the density matrix of an individual state obtained by Eve from the entangling cloner attack. Moreover, $f_{\mathrm{A}}$ is the probability density function of the random variable $A$ representing Alice's random choice for $k_{i}$. Additionally, $f_{\mathrm{A \vert C}}$ is the probability density function of a random variable $A \vert C$ represen\-ting Alice's random choice for $k_{i}$ conditioned on the value of $C$. Note that we use $f_{\mathrm{A\vert C}} = f_{\mathrm{A}}$, since Alice uses the same random variable to get $k_{i}$ independently of the value taken by $C$.
From this description, we write Eve's Holevo quantity for the DR case as
\begin{equation}
\label{Holevo_definition}
\chi_{\mathrm{E,DR}} = S\left(\hat{\rho}_{\mathrm{avg,E}}\right) - \sum_{\mathrm{c = 0,1}} \frac{1}{2}\int_{-\infty}^{\infty}f_{\mathrm{A}}\left(k_{i}\right)\,S\left(\hat{\rho}_{\mathrm{E}}^{k_{i}}\right)\, \mathrm{d}k_{i} \text{,}
\end{equation}
where $S$ is the von Neumann entropy. In order to compute $\chi_{\mathrm{E,RR}}$, we need to compute the covariance matrix of Eve's mode after Bob has performed his measurement on either the \textit{q} or \textit{p} quadrature. Following Ref.\,\citenum{Weedbrook2012}, the covariance matrix of each individual mode of Eve after Bob's measurement is derived as 
\begin{equation}
\label{conditional_covariance_matrix_Eve}
\textbf{V}_{\mathrm{E,B}}^{k_{i}} = \textbf{V}_{\mathrm{avg,E}} - \frac{1}{\sigma_{\mathrm{B}}^2}\textbf{C}_{\mathrm{EB}} \cdot \mathbf{\Pi} \cdot \textbf{C}_{\mathrm{EB}}^{\text{T}} \text{,}
\end{equation}
where $\sigma_{\mathrm{B}}^2 = \tau_{\mathrm{E}}\, e^{2r}/4 + \bar{n} + \bar{n}_{\mathrm{g}} + \left(1-\tau_{\mathrm{E}}\right) /4$. Additionally, $\mathbf{\Pi} \in \{\mathbf{\Pi}_{q}, \mathbf{\Pi}_{p}\}$ is a projective measurement operator in phase space, meaning that
\begin{equation}
\begin{aligned}
\label{V_B}
\mathbf{\Pi}_{q} &= \begin{pmatrix} 1 & 0 \\ 
0 & 0 \end{pmatrix} \, \left(\textit{q} \text{-quadrature measured}\right) \, \text{,} \\
\mathbf{\Pi}_{p} &= \begin{pmatrix} 0 & 0 \\ 
0 & 1 \end{pmatrix} \, \left(\textit{p} \text{-quadrature measured}\right)\text{.}
\end{aligned}
\end{equation}
Finally, $\textbf{C}_{\mathrm{EB}}$ represents the correlations between Eve's mode, which she used during her entangling cloner attack, and Bob's mode. One can derive that 
\begin{equation}
\textbf{C}_{\mathrm{EB}} = \begin{pmatrix}
C_{\mathrm{1}}\,\textbf{I}_{\mathrm{2}}\text{,}\>C_{\mathrm{2}}\,\boldsymbol\sigma_{\mathrm{z}}
\end{pmatrix}^{\mathrm{T}} \text{,}
\end{equation}
where $\boldsymbol\sigma_{\mathrm{z}}$ is the $Z$ Pauli matrix, and 
\begin{equation}
\begin{gathered}
C_{\mathrm{1}} = -\sqrt{\tau_{\mathrm{E}}}\sqrt{1-\tau_{\mathrm{E}}}\left[ e^{2r}/4 - \bar{n}_{\mathrm{tot}}\right] \text{,} \\
C_{\mathrm{2}} = \sqrt{1-\tau_{\mathrm{E}}}\sqrt{\left(\bar{n}_{\mathrm{tot}}\right)^2 -1}\text{.}
\end{gathered}
\end{equation}
In the previous expression, we used the notation $\bar{n}_{\mathrm{tot}} = \left(\bar{n} + \bar{n}_{\mathrm{g}}\right)/\left(1-\tau_{\mathrm{E}}\right) + 1/4$.
Finally, for the RR case one can express Eve's Holevo quantity as
\begin{equation}
\label{Holevo_definition_RR}
\chi_{\mathrm{E,RR}} = S\left(\hat{\rho}_{\mathrm{avg,E}}\right) - \sum_{\mathrm{c = 0,1}} \frac{1}{2}\int_{-\infty}^{\infty}f_{\mathrm{A}}\left(k_{i}\right)\,S\left(\hat{\rho}_{\mathrm{E,B}}^{k_{i}}\right)\, \mathrm{d}k_{i} \text{.}
\end{equation}
Here, $\hat{\rho}_{\mathrm{E,B}}^{k_{i}}$ is the density matrix of Eve's individual mode, which she gets after Bob's individual measurement. The corresponding covariance matrix of Eve's individual mode is then given by Eq.\,\ref{conditional_covariance_matrix_Eve}.
\cite{Mikel18}

\section*{Acknowledgements}
\noindent
We acknowledge support by the German Research Foundation via Germany's Excellence Strategy (EXC-2111-390814868), the Elite Network of Bavaria through the program ExQM, the EU Flagship project QMiCS (Grant No. 820505), and the German Federal Ministry of Education and Research via the project QUARATE (Grant No. 13N15380) and the project QuaMToMe (Grant No. 16KISQ036). This research is part of the Munich Quantum Valley, which is supported by the Bavarian state government with funds from the Hightech Agenda Bayern Plus. 

\noindent We acknowledge helpful discussions with members of the EU Quantum Flagship project Qombs, Alessandro Zavatta, Natalia Bruno, Nicola Biagi, Simone Borri, and Francesco Cappelli.

\section*{Author contributions}
\noindent
K.G.F. and F.D. suggested the idea of the paper. F.F., and K.G.F. developed the initial theory. F.K. and M.R. helped to the development of the final theory. Q.C., Y.N., K.H., O.G., contributed to sections of the manuscript dealing with experimental implementations. K.G.F., A.M., and R.G. supervised this work. F.F., and K.G.F. wrote the manuscript. All authors further contributed to discussions and proofreading of the manuscript.

\section*{Competing interests}
\noindent
The authors declare no competing interests.

\section*{Data availability}
\noindent
Numerical data and codes that support the findings of this study are available from the corresponding author upon reasonable request.

\def\bibsection{\section*{\refname}} 
\bibliography{Perspectives_QKD_open_air_FF}

\begin{thebibliography}{59}%
\makeatletter
\providecommand \@ifxundefined [1]{%
 \@ifx{#1\undefined}
}%
\providecommand \@ifnum [1]{%
 \ifnum #1\expandafter \@firstoftwo
 \else \expandafter \@secondoftwo
 \fi
}%
\providecommand \@ifx [1]{%
 \ifx #1\expandafter \@firstoftwo
 \else \expandafter \@secondoftwo
 \fi
}%
\providecommand \natexlab [1]{#1}%
\providecommand \enquote  [1]{``#1''}%
\providecommand \bibnamefont  [1]{#1}%
\providecommand \bibfnamefont [1]{#1}%
\providecommand \citenamefont [1]{#1}%
\providecommand \href@noop [0]{\@secondoftwo}%
\providecommand \href [0]{\begingroup \@sanitize@url \@href}%
\providecommand \@href[1]{\@@startlink{#1}\@@href}%
\providecommand \@@href[1]{\endgroup#1\@@endlink}%
\providecommand \@sanitize@url [0]{\catcode `\\12\catcode `\$12\catcode
  `\&12\catcode `\#12\catcode `\^12\catcode `\_12\catcode `\%12\relax}%
\providecommand \@@startlink[1]{}%
\providecommand \@@endlink[0]{}%
\providecommand \url  [0]{\begingroup\@sanitize@url \@url }%
\providecommand \@url [1]{\endgroup\@href {#1}{\urlprefix }}%
\providecommand \urlprefix  [0]{URL }%
\providecommand \Eprint [0]{\href }%
\providecommand \doibase [0]{https://doi.org/}%
\providecommand \selectlanguage [0]{\@gobble}%
\providecommand \bibinfo  [0]{\@secondoftwo}%
\providecommand \bibfield  [0]{\@secondoftwo}%
\providecommand \translation [1]{[#1]}%
\providecommand \BibitemOpen [0]{}%
\providecommand \bibitemStop [0]{}%
\providecommand \bibitemNoStop [0]{.\EOS\space}%
\providecommand \EOS [0]{\spacefactor3000\relax}%
\providecommand \BibitemShut  [1]{\csname bibitem#1\endcsname}%
\let\auto@bib@innerbib\@empty
%</preamble>
\bibitem [{\citenamefont {Comandar}\ \emph {et~al.}(2014)\citenamefont
  {Comandar}, \citenamefont {Fr{\"{o}}hlich}, \citenamefont {Lucamarini},
  \citenamefont {Patel}, \citenamefont {Sharpe}, \citenamefont {Dynes},
  \citenamefont {Yuan}, \citenamefont {Penty},\ and\ \citenamefont
  {Shields}}]{Comandar2014}%
  \BibitemOpen
  \bibfield  {author} {\bibinfo {author} {\bibfnamefont {L.~C.}\ \bibnamefont
  {Comandar}}, \bibinfo {author} {\bibfnamefont {B.}~\bibnamefont
  {Fr{\"{o}}hlich}}, \bibinfo {author} {\bibfnamefont {M.}~\bibnamefont
  {Lucamarini}}, \bibinfo {author} {\bibfnamefont {K.~A.}\ \bibnamefont
  {Patel}}, \bibinfo {author} {\bibfnamefont {A.~W.}\ \bibnamefont {Sharpe}},
  \bibinfo {author} {\bibfnamefont {J.~F.}\ \bibnamefont {Dynes}}, \bibinfo
  {author} {\bibfnamefont {Z.~L.}\ \bibnamefont {Yuan}}, \bibinfo {author}
  {\bibfnamefont {R.~V.}\ \bibnamefont {Penty}},\ and\ \bibinfo {author}
  {\bibfnamefont {A.~J.}\ \bibnamefont {Shields}},\ }\bibfield  {title}
  {\bibinfo {title} {{Room temperature single-photon detectors for high bit
  rate quantum key distribution}},\ }\href {https://doi.org/10.1063/1.4855515}
  {\bibfield  {journal} {\bibinfo  {journal} {Applied Physics Letters}\
  }\textbf {\bibinfo {volume} {104}},\ \bibinfo {pages} {021101} (\bibinfo
  {year} {2014})}\BibitemShut {NoStop}%
\bibitem [{\citenamefont {Wang}\ \emph {et~al.}(2018)\citenamefont {Wang},
  \citenamefont {Huang}, \citenamefont {Zhou}, \citenamefont {Liu},
  \citenamefont {Ma}, \citenamefont {Wang},\ and\ \citenamefont
  {Zeng}}]{Wang2018}%
  \BibitemOpen
  \bibfield  {author} {\bibinfo {author} {\bibfnamefont {T.}~\bibnamefont
  {Wang}}, \bibinfo {author} {\bibfnamefont {P.}~\bibnamefont {Huang}},
  \bibinfo {author} {\bibfnamefont {Y.}~\bibnamefont {Zhou}}, \bibinfo {author}
  {\bibfnamefont {W.}~\bibnamefont {Liu}}, \bibinfo {author} {\bibfnamefont
  {H.}~\bibnamefont {Ma}}, \bibinfo {author} {\bibfnamefont {S.}~\bibnamefont
  {Wang}},\ and\ \bibinfo {author} {\bibfnamefont {G.}~\bibnamefont {Zeng}},\
  }\bibfield  {title} {\bibinfo {title} {{High key rate continuous-variable
  quantum key distribution with a real local oscillator}},\ }\href
  {https://doi.org/10.1364/OE.26.002794} {\bibfield  {journal} {\bibinfo
  {journal} {Opt. Express}\ }\textbf {\bibinfo {volume} {26}},\ \bibinfo
  {pages} {2794} (\bibinfo {year} {2018})}\BibitemShut {NoStop}%
\bibitem [{\citenamefont {Diffie}\ and\ \citenamefont
  {Hellman}(1976)}]{DiffieHellman1976}%
  \BibitemOpen
  \bibfield  {author} {\bibinfo {author} {\bibfnamefont {W.}~\bibnamefont
  {Diffie}}\ and\ \bibinfo {author} {\bibfnamefont {M.}~\bibnamefont
  {Hellman}},\ }\bibfield  {title} {\bibinfo {title} {New directions in
  cryptography},\ }\href {https://doi.org/10.1109/TIT.1976.1055638} {\bibfield
  {journal} {\bibinfo  {journal} {IEEE Trans. Inf. Theory}\ }\textbf {\bibinfo
  {volume} {22}},\ \bibinfo {pages} {644} (\bibinfo {year} {1976})}\BibitemShut
  {NoStop}%
\bibitem [{\citenamefont {Rivest}\ \emph {et~al.}(1978)\citenamefont {Rivest},
  \citenamefont {Shamir},\ and\ \citenamefont {Adleman}}]{RSA1978}%
  \BibitemOpen
  \bibfield  {author} {\bibinfo {author} {\bibfnamefont {R.~L.}\ \bibnamefont
  {Rivest}}, \bibinfo {author} {\bibfnamefont {A.}~\bibnamefont {Shamir}},\
  and\ \bibinfo {author} {\bibfnamefont {L.}~\bibnamefont {Adleman}},\
  }\bibfield  {title} {\bibinfo {title} {A method for obtaining digital
  signatures and public-key cryptosystems},\ }\href
  {https://doi.org/10.1145/359340.359342} {\bibfield  {journal} {\bibinfo
  {journal} {Commun. ACM}\ }\textbf {\bibinfo {volume} {21}},\ \bibinfo {pages}
  {120–126} (\bibinfo {year} {1978})}\BibitemShut {NoStop}%
\bibitem [{\citenamefont {Scarani}\ \emph {et~al.}(2009)\citenamefont
  {Scarani}, \citenamefont {Bechmann-Pasquinucci}, \citenamefont {Cerf},
  \citenamefont {Du{\v{s}}ek}, \citenamefont {L{\"{u}}tkenhaus},\ and\
  \citenamefont {Peev}}]{Scarani2009}%
  \BibitemOpen
  \bibfield  {author} {\bibinfo {author} {\bibfnamefont {V.}~\bibnamefont
  {Scarani}}, \bibinfo {author} {\bibfnamefont {H.}~\bibnamefont
  {Bechmann-Pasquinucci}}, \bibinfo {author} {\bibfnamefont {N.~J.}\
  \bibnamefont {Cerf}}, \bibinfo {author} {\bibfnamefont {M.}~\bibnamefont
  {Du{\v{s}}ek}}, \bibinfo {author} {\bibfnamefont {N.}~\bibnamefont
  {L{\"{u}}tkenhaus}},\ and\ \bibinfo {author} {\bibfnamefont {M.}~\bibnamefont
  {Peev}},\ }\bibfield  {title} {\bibinfo {title} {{The security of practical
  quantum key distribution}},\ }\href
  {https://doi.org/10.1103/RevModPhys.81.1301} {\bibfield  {journal} {\bibinfo
  {journal} {Rev. Mod. Phys}\ }\textbf {\bibinfo {volume} {81}},\ \bibinfo
  {pages} {1301} (\bibinfo {year} {2009})}\BibitemShut {NoStop}%
\bibitem [{\citenamefont {Zurek}(1973)}]{Zurek1973}%
  \BibitemOpen
  \bibfield  {author} {\bibinfo {author} {\bibfnamefont {W.~H.}\ \bibnamefont
  {Zurek}},\ }\bibfield  {title} {\bibinfo {title} {{A single quantum cannot be
  cloned}},\ }\href {https://doi.org/10.1038/246170a0} {\bibfield  {journal}
  {\bibinfo  {journal} {Nature}\ }\textbf {\bibinfo {volume} {246}},\ \bibinfo
  {pages} {170} (\bibinfo {year} {1973})}\BibitemShut {NoStop}%
\bibitem [{\citenamefont {Laudenbach}\ \emph {et~al.}(2018)\citenamefont
  {Laudenbach}, \citenamefont {Pacher}, \citenamefont {Fung}, \citenamefont
  {Poppe}, \citenamefont {Peev}, \citenamefont {Schrenk}, \citenamefont
  {Hentschel}, \citenamefont {Walther},\ and\ \citenamefont
  {H{\"{u}}bel}}]{Laudenbach2018}%
  \BibitemOpen
  \bibfield  {author} {\bibinfo {author} {\bibfnamefont {F.}~\bibnamefont
  {Laudenbach}}, \bibinfo {author} {\bibfnamefont {C.}~\bibnamefont {Pacher}},
  \bibinfo {author} {\bibfnamefont {C.-H.~F.}\ \bibnamefont {Fung}}, \bibinfo
  {author} {\bibfnamefont {A.}~\bibnamefont {Poppe}}, \bibinfo {author}
  {\bibfnamefont {M.}~\bibnamefont {Peev}}, \bibinfo {author} {\bibfnamefont
  {B.}~\bibnamefont {Schrenk}}, \bibinfo {author} {\bibfnamefont
  {M.}~\bibnamefont {Hentschel}}, \bibinfo {author} {\bibfnamefont
  {P.}~\bibnamefont {Walther}},\ and\ \bibinfo {author} {\bibfnamefont
  {H.}~\bibnamefont {H{\"{u}}bel}},\ }\bibfield  {title} {\bibinfo {title}
  {{Continuous-Variable Quantum Key Distribution with Gaussian Modulation-The
  Theory of Practical Implementations}},\ }\href
  {https://doi.org/10.1002/qute.201800011} {\bibfield  {journal} {\bibinfo
  {journal} {Adv. Quantum Technol.}\ }\textbf {\bibinfo {volume} {1}},\
  \bibinfo {pages} {1800011} (\bibinfo {year} {2018})}\BibitemShut {NoStop}%
\bibitem [{\citenamefont {Lodewyck}\ \emph {et~al.}(2005)\citenamefont
  {Lodewyck}, \citenamefont {Debuisschert}, \citenamefont {Tualle-Brouri},\
  and\ \citenamefont {Grangier}}]{Lodewyck2005}%
  \BibitemOpen
  \bibfield  {author} {\bibinfo {author} {\bibfnamefont {J.}~\bibnamefont
  {Lodewyck}}, \bibinfo {author} {\bibfnamefont {T.}~\bibnamefont
  {Debuisschert}}, \bibinfo {author} {\bibfnamefont {R.}~\bibnamefont
  {Tualle-Brouri}},\ and\ \bibinfo {author} {\bibfnamefont {P.}~\bibnamefont
  {Grangier}},\ }\bibfield  {title} {\bibinfo {title} {{Controlling excess
  noise in fiber-optics continuous-variable quantum key distribution}},\ }\href
  {https://doi.org/10.1103/PhysRevA.72.050303} {\bibfield  {journal} {\bibinfo
  {journal} {Phys. Rev. A}\ }\textbf {\bibinfo {volume} {72}},\ \bibinfo
  {pages} {050303} (\bibinfo {year} {2005})}\BibitemShut {NoStop}%
\bibitem [{\citenamefont {Grosshans}\ \emph
  {et~al.}(2003{\natexlab{a}})\citenamefont {Grosshans}, \citenamefont
  {Van~Assche}, \citenamefont {Wenger}, \citenamefont {Brouri}, \citenamefont
  {Cerf},\ and\ \citenamefont {Grangier}}]{Grosshans2003}%
  \BibitemOpen
  \bibfield  {author} {\bibinfo {author} {\bibfnamefont {F.}~\bibnamefont
  {Grosshans}}, \bibinfo {author} {\bibfnamefont {G.}~\bibnamefont
  {Van~Assche}}, \bibinfo {author} {\bibfnamefont {J.}~\bibnamefont {Wenger}},
  \bibinfo {author} {\bibfnamefont {R.}~\bibnamefont {Brouri}}, \bibinfo
  {author} {\bibfnamefont {N.~J.}\ \bibnamefont {Cerf}},\ and\ \bibinfo
  {author} {\bibfnamefont {P.}~\bibnamefont {Grangier}},\ }\bibfield  {title}
  {\bibinfo {title} {Quantum key distribution using gaussian-modulated coherent
  states},\ }\href {https://doi.org/10.1038/nature01289} {\bibfield  {journal}
  {\bibinfo  {journal} {Nature}\ }\textbf {\bibinfo {volume} {421}},\ \bibinfo
  {pages} {238} (\bibinfo {year} {2003}{\natexlab{a}})}\BibitemShut {NoStop}%
\bibitem [{\citenamefont {Buttler}\ \emph {et~al.}(2000)\citenamefont
  {Buttler}, \citenamefont {Hughes}, \citenamefont {Lamoreaux}, \citenamefont
  {Morgan}, \citenamefont {Nordholt},\ and\ \citenamefont
  {Peterson}}]{Buttler2000}%
  \BibitemOpen
  \bibfield  {author} {\bibinfo {author} {\bibfnamefont {W.~T.}\ \bibnamefont
  {Buttler}}, \bibinfo {author} {\bibfnamefont {R.~J.}\ \bibnamefont {Hughes}},
  \bibinfo {author} {\bibfnamefont {S.~K.}\ \bibnamefont {Lamoreaux}}, \bibinfo
  {author} {\bibfnamefont {G.~L.}\ \bibnamefont {Morgan}}, \bibinfo {author}
  {\bibfnamefont {J.~E.}\ \bibnamefont {Nordholt}},\ and\ \bibinfo {author}
  {\bibfnamefont {C.~G.}\ \bibnamefont {Peterson}},\ }\bibfield  {title}
  {\bibinfo {title} {Daylight quantum key distribution over 1.6 km},\ }\href
  {https://doi.org/10.1103/PhysRevLett.84.5652} {\bibfield  {journal} {\bibinfo
   {journal} {Phys. Rev. Lett.}\ }\textbf {\bibinfo {volume} {84}},\ \bibinfo
  {pages} {5652} (\bibinfo {year} {2000})}\BibitemShut {NoStop}%
\bibitem [{\citenamefont {Liao}\ \emph {et~al.}(2017)\citenamefont {Liao},
  \citenamefont {Cai}, \citenamefont {Liu}, \citenamefont {Zhang},
  \citenamefont {Li}, \citenamefont {Ren}, \citenamefont {Yin}, \citenamefont
  {Shen}, \citenamefont {Cao}, \citenamefont {Li}, \citenamefont {Li},
  \citenamefont {Chen}, \citenamefont {Sun}, \citenamefont {Jia}, \citenamefont
  {Wu}, \citenamefont {Jiang}, \citenamefont {Wang}, \citenamefont {Huang},
  \citenamefont {Wang}, \citenamefont {Zhou}, \citenamefont {Deng},
  \citenamefont {Xi}, \citenamefont {Ma}, \citenamefont {Hu}, \citenamefont
  {Zhang}, \citenamefont {Chen}, \citenamefont {Liu}, \citenamefont {Wang},
  \citenamefont {Zhu}, \citenamefont {Lu}, \citenamefont {Shu}, \citenamefont
  {Peng}, \citenamefont {Wang},\ and\ \citenamefont {Pan}}]{Liao2017}%
  \BibitemOpen
  \bibfield  {author} {\bibinfo {author} {\bibfnamefont {S.-K.}\ \bibnamefont
  {Liao}}, \bibinfo {author} {\bibfnamefont {W.-Q.}\ \bibnamefont {Cai}},
  \bibinfo {author} {\bibfnamefont {W.-Y.}\ \bibnamefont {Liu}}, \bibinfo
  {author} {\bibfnamefont {L.}~\bibnamefont {Zhang}}, \bibinfo {author}
  {\bibfnamefont {Y.}~\bibnamefont {Li}}, \bibinfo {author} {\bibfnamefont
  {J.-G.}\ \bibnamefont {Ren}}, \bibinfo {author} {\bibfnamefont
  {J.}~\bibnamefont {Yin}}, \bibinfo {author} {\bibfnamefont {Q.}~\bibnamefont
  {Shen}}, \bibinfo {author} {\bibfnamefont {Y.}~\bibnamefont {Cao}}, \bibinfo
  {author} {\bibfnamefont {Z.-P.}\ \bibnamefont {Li}}, \bibinfo {author}
  {\bibfnamefont {F.-Z.}\ \bibnamefont {Li}}, \bibinfo {author} {\bibfnamefont
  {X.-W.}\ \bibnamefont {Chen}}, \bibinfo {author} {\bibfnamefont {L.-H.}\
  \bibnamefont {Sun}}, \bibinfo {author} {\bibfnamefont {J.-J.}\ \bibnamefont
  {Jia}}, \bibinfo {author} {\bibfnamefont {J.-C.}\ \bibnamefont {Wu}},
  \bibinfo {author} {\bibfnamefont {X.-J.}\ \bibnamefont {Jiang}}, \bibinfo
  {author} {\bibfnamefont {J.-F.}\ \bibnamefont {Wang}}, \bibinfo {author}
  {\bibfnamefont {Y.-M.}\ \bibnamefont {Huang}}, \bibinfo {author}
  {\bibfnamefont {Q.}~\bibnamefont {Wang}}, \bibinfo {author} {\bibfnamefont
  {Y.-L.}\ \bibnamefont {Zhou}}, \bibinfo {author} {\bibfnamefont
  {L.}~\bibnamefont {Deng}}, \bibinfo {author} {\bibfnamefont {T.}~\bibnamefont
  {Xi}}, \bibinfo {author} {\bibfnamefont {L.}~\bibnamefont {Ma}}, \bibinfo
  {author} {\bibfnamefont {T.}~\bibnamefont {Hu}}, \bibinfo {author}
  {\bibfnamefont {Q.}~\bibnamefont {Zhang}}, \bibinfo {author} {\bibfnamefont
  {Y.-A.}\ \bibnamefont {Chen}}, \bibinfo {author} {\bibfnamefont {N.-L.}\
  \bibnamefont {Liu}}, \bibinfo {author} {\bibfnamefont {X.-B.}\ \bibnamefont
  {Wang}}, \bibinfo {author} {\bibfnamefont {Z.-C.}\ \bibnamefont {Zhu}},
  \bibinfo {author} {\bibfnamefont {C.-Y.}\ \bibnamefont {Lu}}, \bibinfo
  {author} {\bibfnamefont {R.}~\bibnamefont {Shu}}, \bibinfo {author}
  {\bibfnamefont {C.-Z.}\ \bibnamefont {Peng}}, \bibinfo {author}
  {\bibfnamefont {J.-Y.}\ \bibnamefont {Wang}},\ and\ \bibinfo {author}
  {\bibfnamefont {J.-W.}\ \bibnamefont {Pan}},\ }\bibfield  {title} {\bibinfo
  {title} {Satellite-to-ground quantum key distribution},\ }\href
  {https://doi.org/10.1038/nature23655} {\bibfield  {journal} {\bibinfo
  {journal} {Nature}\ }\textbf {\bibinfo {volume} {549}},\ \bibinfo {pages}
  {43} (\bibinfo {year} {2017})}\BibitemShut {NoStop}%
\bibitem [{\citenamefont {Kaushal}\ \emph {et~al.}(2018)\citenamefont
  {Kaushal}, \citenamefont {Jain},\ and\ \citenamefont {Kar}}]{Kaushal2018}%
  \BibitemOpen
  \bibfield  {author} {\bibinfo {author} {\bibfnamefont {H.}~\bibnamefont
  {Kaushal}}, \bibinfo {author} {\bibfnamefont {V.}~\bibnamefont {Jain}},\ and\
  \bibinfo {author} {\bibfnamefont {S.}~\bibnamefont {Kar}},\ }\href
  {https://doi.org/10.1007/978-81-322-3691-7} {\emph {\bibinfo {title} {{Free
  Space Optical Communication}}}},\ Vol.~\bibinfo {volume} {7}\ (\bibinfo
  {publisher} {Springer},\ \bibinfo {address} {New Delhi},\ \bibinfo {year}
  {2018})\BibitemShut {NoStop}%
\bibitem [{\citenamefont {Arute}\ \emph {et~al.}(2019)\citenamefont {Arute},
  \citenamefont {Arya}, \citenamefont {Babbush}, \citenamefont {Bacon},
  \citenamefont {Bardin}, \citenamefont {Barends}, \citenamefont {Biswas},
  \citenamefont {Boixo}, \citenamefont {Brandao}, \citenamefont {Buell},
  \citenamefont {Burkett}, \citenamefont {Chen}, \citenamefont {Chen},
  \citenamefont {Chiaro}, \citenamefont {Collins}, \citenamefont {Courtney},
  \citenamefont {Dunsworth}, \citenamefont {Farhi}, \citenamefont {Foxen},
  \citenamefont {Fowler}, \citenamefont {Gidney}, \citenamefont {Giustina},
  \citenamefont {Graff}, \citenamefont {Guerin}, \citenamefont {Habegger},
  \citenamefont {Harrigan}, \citenamefont {Hartmann}, \citenamefont {Ho},
  \citenamefont {Hoffmann}, \citenamefont {Huang}, \citenamefont {Humble},
  \citenamefont {Isakov}, \citenamefont {Jeffrey}, \citenamefont {Jiang},
  \citenamefont {Kafri}, \citenamefont {Kechedzhi}, \citenamefont {Kelly},
  \citenamefont {Klimov}, \citenamefont {Knysh}, \citenamefont {Korotkov},
  \citenamefont {Kostritsa}, \citenamefont {Landhuis}, \citenamefont
  {Lindmark}, \citenamefont {Lucero}, \citenamefont {Lyakh}, \citenamefont
  {Mandr{\`{a}}}, \citenamefont {McClean}, \citenamefont {McEwen},
  \citenamefont {Megrant}, \citenamefont {Mi}, \citenamefont {Michielsen},
  \citenamefont {Mohseni}, \citenamefont {Mutus}, \citenamefont {Naaman},
  \citenamefont {Neeley}, \citenamefont {Neill}, \citenamefont {Niu},
  \citenamefont {Ostby}, \citenamefont {Petukhov}, \citenamefont {Platt},
  \citenamefont {Quintana}, \citenamefont {Rieffel}, \citenamefont {Roushan},
  \citenamefont {Rubin}, \citenamefont {Sank}, \citenamefont {Satzinger},
  \citenamefont {Smelyanskiy}, \citenamefont {Sung}, \citenamefont
  {Trevithick}, \citenamefont {Vainsencher}, \citenamefont {Villalonga},
  \citenamefont {White}, \citenamefont {Yao}, \citenamefont {Yeh},
  \citenamefont {Zalcman}, \citenamefont {Neven},\ and\ \citenamefont
  {Martinis}}]{Arute2019}%
  \BibitemOpen
  \bibfield  {author} {\bibinfo {author} {\bibfnamefont {F.}~\bibnamefont
  {Arute}}, \bibinfo {author} {\bibfnamefont {K.}~\bibnamefont {Arya}},
  \bibinfo {author} {\bibfnamefont {R.}~\bibnamefont {Babbush}}, \bibinfo
  {author} {\bibfnamefont {D.}~\bibnamefont {Bacon}}, \bibinfo {author}
  {\bibfnamefont {J.~C.}\ \bibnamefont {Bardin}}, \bibinfo {author}
  {\bibfnamefont {R.}~\bibnamefont {Barends}}, \bibinfo {author} {\bibfnamefont
  {R.}~\bibnamefont {Biswas}}, \bibinfo {author} {\bibfnamefont
  {S.}~\bibnamefont {Boixo}}, \bibinfo {author} {\bibfnamefont {F.~G. S.~L.}\
  \bibnamefont {Brandao}}, \bibinfo {author} {\bibfnamefont {D.~A.}\
  \bibnamefont {Buell}}, \bibinfo {author} {\bibfnamefont {B.}~\bibnamefont
  {Burkett}}, \bibinfo {author} {\bibfnamefont {Y.}~\bibnamefont {Chen}},
  \bibinfo {author} {\bibfnamefont {Z.}~\bibnamefont {Chen}}, \bibinfo {author}
  {\bibfnamefont {B.}~\bibnamefont {Chiaro}}, \bibinfo {author} {\bibfnamefont
  {R.}~\bibnamefont {Collins}}, \bibinfo {author} {\bibfnamefont
  {W.}~\bibnamefont {Courtney}}, \bibinfo {author} {\bibfnamefont
  {A.}~\bibnamefont {Dunsworth}}, \bibinfo {author} {\bibfnamefont
  {E.}~\bibnamefont {Farhi}}, \bibinfo {author} {\bibfnamefont
  {B.}~\bibnamefont {Foxen}}, \bibinfo {author} {\bibfnamefont
  {A.}~\bibnamefont {Fowler}}, \bibinfo {author} {\bibfnamefont
  {C.}~\bibnamefont {Gidney}}, \bibinfo {author} {\bibfnamefont
  {M.}~\bibnamefont {Giustina}}, \bibinfo {author} {\bibfnamefont
  {R.}~\bibnamefont {Graff}}, \bibinfo {author} {\bibfnamefont
  {K.}~\bibnamefont {Guerin}}, \bibinfo {author} {\bibfnamefont
  {S.}~\bibnamefont {Habegger}}, \bibinfo {author} {\bibfnamefont {M.~P.}\
  \bibnamefont {Harrigan}}, \bibinfo {author} {\bibfnamefont {M.~J.}\
  \bibnamefont {Hartmann}}, \bibinfo {author} {\bibfnamefont {A.}~\bibnamefont
  {Ho}}, \bibinfo {author} {\bibfnamefont {M.}~\bibnamefont {Hoffmann}},
  \bibinfo {author} {\bibfnamefont {T.}~\bibnamefont {Huang}}, \bibinfo
  {author} {\bibfnamefont {T.~S.}\ \bibnamefont {Humble}}, \bibinfo {author}
  {\bibfnamefont {S.~V.}\ \bibnamefont {Isakov}}, \bibinfo {author}
  {\bibfnamefont {E.}~\bibnamefont {Jeffrey}}, \bibinfo {author} {\bibfnamefont
  {Z.}~\bibnamefont {Jiang}}, \bibinfo {author} {\bibfnamefont
  {D.}~\bibnamefont {Kafri}}, \bibinfo {author} {\bibfnamefont
  {K.}~\bibnamefont {Kechedzhi}}, \bibinfo {author} {\bibfnamefont
  {J.}~\bibnamefont {Kelly}}, \bibinfo {author} {\bibfnamefont {P.~V.}\
  \bibnamefont {Klimov}}, \bibinfo {author} {\bibfnamefont {S.}~\bibnamefont
  {Knysh}}, \bibinfo {author} {\bibfnamefont {A.}~\bibnamefont {Korotkov}},
  \bibinfo {author} {\bibfnamefont {F.}~\bibnamefont {Kostritsa}}, \bibinfo
  {author} {\bibfnamefont {D.}~\bibnamefont {Landhuis}}, \bibinfo {author}
  {\bibfnamefont {M.}~\bibnamefont {Lindmark}}, \bibinfo {author}
  {\bibfnamefont {E.}~\bibnamefont {Lucero}}, \bibinfo {author} {\bibfnamefont
  {D.}~\bibnamefont {Lyakh}}, \bibinfo {author} {\bibfnamefont
  {S.}~\bibnamefont {Mandr{\`{a}}}}, \bibinfo {author} {\bibfnamefont {J.~R.}\
  \bibnamefont {McClean}}, \bibinfo {author} {\bibfnamefont {M.}~\bibnamefont
  {McEwen}}, \bibinfo {author} {\bibfnamefont {A.}~\bibnamefont {Megrant}},
  \bibinfo {author} {\bibfnamefont {X.}~\bibnamefont {Mi}}, \bibinfo {author}
  {\bibfnamefont {K.}~\bibnamefont {Michielsen}}, \bibinfo {author}
  {\bibfnamefont {M.}~\bibnamefont {Mohseni}}, \bibinfo {author} {\bibfnamefont
  {J.}~\bibnamefont {Mutus}}, \bibinfo {author} {\bibfnamefont
  {O.}~\bibnamefont {Naaman}}, \bibinfo {author} {\bibfnamefont
  {M.}~\bibnamefont {Neeley}}, \bibinfo {author} {\bibfnamefont
  {C.}~\bibnamefont {Neill}}, \bibinfo {author} {\bibfnamefont {M.~Y.}\
  \bibnamefont {Niu}}, \bibinfo {author} {\bibfnamefont {E.}~\bibnamefont
  {Ostby}}, \bibinfo {author} {\bibfnamefont {A.}~\bibnamefont {Petukhov}},
  \bibinfo {author} {\bibfnamefont {J.~C.}\ \bibnamefont {Platt}}, \bibinfo
  {author} {\bibfnamefont {C.}~\bibnamefont {Quintana}}, \bibinfo {author}
  {\bibfnamefont {E.~G.}\ \bibnamefont {Rieffel}}, \bibinfo {author}
  {\bibfnamefont {P.}~\bibnamefont {Roushan}}, \bibinfo {author} {\bibfnamefont
  {N.~C.}\ \bibnamefont {Rubin}}, \bibinfo {author} {\bibfnamefont
  {D.}~\bibnamefont {Sank}}, \bibinfo {author} {\bibfnamefont {K.~J.}\
  \bibnamefont {Satzinger}}, \bibinfo {author} {\bibfnamefont {V.}~\bibnamefont
  {Smelyanskiy}}, \bibinfo {author} {\bibfnamefont {K.~J.}\ \bibnamefont
  {Sung}}, \bibinfo {author} {\bibfnamefont {M.~D.}\ \bibnamefont
  {Trevithick}}, \bibinfo {author} {\bibfnamefont {A.}~\bibnamefont
  {Vainsencher}}, \bibinfo {author} {\bibfnamefont {B.}~\bibnamefont
  {Villalonga}}, \bibinfo {author} {\bibfnamefont {T.}~\bibnamefont {White}},
  \bibinfo {author} {\bibfnamefont {Z.~J.}\ \bibnamefont {Yao}}, \bibinfo
  {author} {\bibfnamefont {P.}~\bibnamefont {Yeh}}, \bibinfo {author}
  {\bibfnamefont {A.}~\bibnamefont {Zalcman}}, \bibinfo {author} {\bibfnamefont
  {H.}~\bibnamefont {Neven}},\ and\ \bibinfo {author} {\bibfnamefont {J.~M.}\
  \bibnamefont {Martinis}},\ }\bibfield  {title} {\bibinfo {title} {{Quantum
  supremacy using a programmable superconducting processor}},\ }\href
  {https://doi.org/10.1038/s41586-019-1666-5} {\bibfield  {journal} {\bibinfo
  {journal} {Nature}\ }\textbf {\bibinfo {volume} {574}},\ \bibinfo {pages}
  {505} (\bibinfo {year} {2019})}\BibitemShut {NoStop}%
\bibitem [{\citenamefont {Kjaergaard}\ \emph {et~al.}(2020)\citenamefont
  {Kjaergaard}, \citenamefont {Schwartz}, \citenamefont {Braum{\"{u}}ller},
  \citenamefont {Krantz}, \citenamefont {Wang}, \citenamefont {Gustavsson},\
  and\ \citenamefont {Oliver}}]{Kjaergaard2020}%
  \BibitemOpen
  \bibfield  {author} {\bibinfo {author} {\bibfnamefont {M.}~\bibnamefont
  {Kjaergaard}}, \bibinfo {author} {\bibfnamefont {M.~E.}\ \bibnamefont
  {Schwartz}}, \bibinfo {author} {\bibfnamefont {J.}~\bibnamefont
  {Braum{\"{u}}ller}}, \bibinfo {author} {\bibfnamefont {P.}~\bibnamefont
  {Krantz}}, \bibinfo {author} {\bibfnamefont {J.~I.}\ \bibnamefont {Wang}},
  \bibinfo {author} {\bibfnamefont {S.}~\bibnamefont {Gustavsson}},\ and\
  \bibinfo {author} {\bibfnamefont {W.~D.}\ \bibnamefont {Oliver}},\ }\bibfield
   {title} {\bibinfo {title} {{Superconducting Qubits: Current State of
  Play}},\ }\href {https://doi.org/10.1146/annurev-conmatphys-031119-050605}
  {\bibfield  {journal} {\bibinfo  {journal} {Annu. Rev. Condens. Matter
  Phys.}\ }\textbf {\bibinfo {volume} {11}},\ \bibinfo {pages} {369} (\bibinfo
  {year} {2020})}\BibitemShut {NoStop}%
\bibitem [{\citenamefont {Pogorzalek}\ \emph {et~al.}(2019)\citenamefont
  {Pogorzalek}, \citenamefont {Fedorov}, \citenamefont {Xu}, \citenamefont
  {Parra-Rodriguez}, \citenamefont {Sanz}, \citenamefont {Fischer},
  \citenamefont {Xie}, \citenamefont {Inomata}, \citenamefont {Nakamura},
  \citenamefont {Solano}, \citenamefont {Marx}, \citenamefont {Deppe},\ and\
  \citenamefont {Gross}}]{Pogorzalek2019}%
  \BibitemOpen
  \bibfield  {author} {\bibinfo {author} {\bibfnamefont {S.}~\bibnamefont
  {Pogorzalek}}, \bibinfo {author} {\bibfnamefont {K.~G.}\ \bibnamefont
  {Fedorov}}, \bibinfo {author} {\bibfnamefont {M.}~\bibnamefont {Xu}},
  \bibinfo {author} {\bibfnamefont {A.}~\bibnamefont {Parra-Rodriguez}},
  \bibinfo {author} {\bibfnamefont {M.}~\bibnamefont {Sanz}}, \bibinfo {author}
  {\bibfnamefont {M.}~\bibnamefont {Fischer}}, \bibinfo {author} {\bibfnamefont
  {E.}~\bibnamefont {Xie}}, \bibinfo {author} {\bibfnamefont {K.}~\bibnamefont
  {Inomata}}, \bibinfo {author} {\bibfnamefont {Y.}~\bibnamefont {Nakamura}},
  \bibinfo {author} {\bibfnamefont {E.}~\bibnamefont {Solano}}, \bibinfo
  {author} {\bibfnamefont {A.}~\bibnamefont {Marx}}, \bibinfo {author}
  {\bibfnamefont {F.}~\bibnamefont {Deppe}},\ and\ \bibinfo {author}
  {\bibfnamefont {R.}~\bibnamefont {Gross}},\ }\bibfield  {title} {\bibinfo
  {title} {{Secure quantum remote state preparation of squeezed microwave
  states}},\ }\href {https://doi.org/10.1038/s41467-019-10727-7} {\bibfield
  {journal} {\bibinfo  {journal} {Nat. Commun.}\ }\textbf {\bibinfo {volume}
  {10}},\ \bibinfo {pages} {2604} (\bibinfo {year} {2019})}\BibitemShut
  {NoStop}%
\bibitem [{\citenamefont {Bienfait}\ \emph {et~al.}(2019)\citenamefont
  {Bienfait}, \citenamefont {Satzinger}, \citenamefont {Zhong}, \citenamefont
  {Chang}, \citenamefont {Chou}, \citenamefont {Conner}, \citenamefont {Dumur},
  \citenamefont {Grebel}, \citenamefont {Peairs}, \citenamefont {Povey},\ and\
  \citenamefont {Cleland}}]{Bienfait2019}%
  \BibitemOpen
  \bibfield  {author} {\bibinfo {author} {\bibfnamefont {A.}~\bibnamefont
  {Bienfait}}, \bibinfo {author} {\bibfnamefont {K.~J.}\ \bibnamefont
  {Satzinger}}, \bibinfo {author} {\bibfnamefont {Y.~P.}\ \bibnamefont
  {Zhong}}, \bibinfo {author} {\bibfnamefont {H.-S.}\ \bibnamefont {Chang}},
  \bibinfo {author} {\bibfnamefont {M.-H.}\ \bibnamefont {Chou}}, \bibinfo
  {author} {\bibfnamefont {C.~R.}\ \bibnamefont {Conner}}, \bibinfo {author}
  {\bibfnamefont {{\'{E}}.}~\bibnamefont {Dumur}}, \bibinfo {author}
  {\bibfnamefont {J.}~\bibnamefont {Grebel}}, \bibinfo {author} {\bibfnamefont
  {G.~A.}\ \bibnamefont {Peairs}}, \bibinfo {author} {\bibfnamefont {R.~G.}\
  \bibnamefont {Povey}},\ and\ \bibinfo {author} {\bibfnamefont {A.~N.}\
  \bibnamefont {Cleland}},\ }\bibfield  {title} {\bibinfo {title}
  {{Phonon-mediated quantum state transfer and remote qubit entanglement}},\
  }\href {https://doi.org/10.1126/science.aaw8415} {\bibfield  {journal}
  {\bibinfo  {journal} {Science}\ }\textbf {\bibinfo {volume} {364}},\ \bibinfo
  {pages} {368} (\bibinfo {year} {2019})}\BibitemShut {NoStop}%
\bibitem [{\citenamefont {Zhong}\ \emph {et~al.}(2013)\citenamefont {Zhong},
  \citenamefont {Menzel}, \citenamefont {{Di Candia}}, \citenamefont {Eder},
  \citenamefont {Ihmig}, \citenamefont {Baust}, \citenamefont {Haeberlein},
  \citenamefont {Hoffmann}, \citenamefont {Inomata}, \citenamefont {Yamamoto},
  \citenamefont {Nakamura}, \citenamefont {Solano}, \citenamefont {Deppe},
  \citenamefont {Marx},\ and\ \citenamefont {Gross}}]{Zhong2013}%
  \BibitemOpen
  \bibfield  {author} {\bibinfo {author} {\bibfnamefont {L.}~\bibnamefont
  {Zhong}}, \bibinfo {author} {\bibfnamefont {E.~P.}\ \bibnamefont {Menzel}},
  \bibinfo {author} {\bibfnamefont {R.}~\bibnamefont {{Di Candia}}}, \bibinfo
  {author} {\bibfnamefont {P.}~\bibnamefont {Eder}}, \bibinfo {author}
  {\bibfnamefont {M.}~\bibnamefont {Ihmig}}, \bibinfo {author} {\bibfnamefont
  {A.}~\bibnamefont {Baust}}, \bibinfo {author} {\bibfnamefont
  {M.}~\bibnamefont {Haeberlein}}, \bibinfo {author} {\bibfnamefont
  {E.}~\bibnamefont {Hoffmann}}, \bibinfo {author} {\bibfnamefont
  {K.}~\bibnamefont {Inomata}}, \bibinfo {author} {\bibfnamefont
  {T.}~\bibnamefont {Yamamoto}}, \bibinfo {author} {\bibfnamefont
  {Y.}~\bibnamefont {Nakamura}}, \bibinfo {author} {\bibfnamefont
  {E.}~\bibnamefont {Solano}}, \bibinfo {author} {\bibfnamefont
  {F.}~\bibnamefont {Deppe}}, \bibinfo {author} {\bibfnamefont
  {A.}~\bibnamefont {Marx}},\ and\ \bibinfo {author} {\bibfnamefont
  {R.}~\bibnamefont {Gross}},\ }\bibfield  {title} {\bibinfo {title}
  {{Squeezing with a flux-driven Josephson parametric amplifier}},\ }\href
  {https://doi.org/10.1088/1367-2630/15/12/125013} {\bibfield  {journal}
  {\bibinfo  {journal} {New J. Phys.}\ }\textbf {\bibinfo {volume} {15}},\
  \bibinfo {pages} {125013} (\bibinfo {year} {2013})}\BibitemShut {NoStop}%
\bibitem [{\citenamefont {Eichler}\ \emph {et~al.}(2014)\citenamefont
  {Eichler}, \citenamefont {Salathe}, \citenamefont {Mlynek}, \citenamefont
  {Schmidt},\ and\ \citenamefont {Wallraff}}]{Eichler2014}%
  \BibitemOpen
  \bibfield  {author} {\bibinfo {author} {\bibfnamefont {C.}~\bibnamefont
  {Eichler}}, \bibinfo {author} {\bibfnamefont {Y.}~\bibnamefont {Salathe}},
  \bibinfo {author} {\bibfnamefont {J.}~\bibnamefont {Mlynek}}, \bibinfo
  {author} {\bibfnamefont {S.}~\bibnamefont {Schmidt}},\ and\ \bibinfo {author}
  {\bibfnamefont {A.}~\bibnamefont {Wallraff}},\ }\bibfield  {title} {\bibinfo
  {title} {{Quantum-Limited Amplification and Entanglement in Coupled Nonlinear
  Resonators}},\ }\href {https://doi.org/10.1103/PhysRevLett.113.110502}
  {\bibfield  {journal} {\bibinfo  {journal} {Phys. Rev. Lett}\ }\textbf
  {\bibinfo {volume} {113}},\ \bibinfo {pages} {110502} (\bibinfo {year}
  {2014})}\BibitemShut {NoStop}%
\bibitem [{\citenamefont {Grimsmo}\ and\ \citenamefont
  {Blais}(2017)}]{Grimsmo2017}%
  \BibitemOpen
  \bibfield  {author} {\bibinfo {author} {\bibfnamefont {A.~L.}\ \bibnamefont
  {Grimsmo}}\ and\ \bibinfo {author} {\bibfnamefont {A.}~\bibnamefont
  {Blais}},\ }\bibfield  {title} {\bibinfo {title} {{Squeezing and quantum
  state engineering with Josephson travelling wave amplifiers}},\ }\href
  {https://doi.org/10.1038/s41534-017-0020-8} {\bibfield  {journal} {\bibinfo
  {journal} {Npj Quantum Inf.}\ }\textbf {\bibinfo {volume} {3}},\ \bibinfo
  {pages} {20} (\bibinfo {year} {2017})}\BibitemShut {NoStop}%
\bibitem [{\citenamefont {Cerf}\ \emph {et~al.}(2001)\citenamefont {Cerf},
  \citenamefont {L\'evy},\ and\ \citenamefont {Assche}}]{Cerf2001}%
  \BibitemOpen
  \bibfield  {author} {\bibinfo {author} {\bibfnamefont {N.~J.}\ \bibnamefont
  {Cerf}}, \bibinfo {author} {\bibfnamefont {M.}~\bibnamefont {L\'evy}},\ and\
  \bibinfo {author} {\bibfnamefont {G.~V.}\ \bibnamefont {Assche}},\ }\bibfield
   {title} {\bibinfo {title} {Quantum distribution of gaussian keys using
  squeezed states},\ }\href {https://doi.org/10.1103/PhysRevA.63.052311}
  {\bibfield  {journal} {\bibinfo  {journal} {Phys. Rev. A}\ }\textbf {\bibinfo
  {volume} {63}},\ \bibinfo {pages} {052311} (\bibinfo {year}
  {2001})}\BibitemShut {NoStop}%
\bibitem [{\citenamefont {Grosshans}\ \emph
  {et~al.}(2003{\natexlab{b}})\citenamefont {Grosshans}, \citenamefont
  {Van~Assche}, \citenamefont {Wenger}, \citenamefont {Brouri}, \citenamefont
  {Cerf},\ and\ \citenamefont {Grangier}}]{grosshans2003_3}%
  \BibitemOpen
  \bibfield  {author} {\bibinfo {author} {\bibfnamefont {F.}~\bibnamefont
  {Grosshans}}, \bibinfo {author} {\bibfnamefont {G.}~\bibnamefont
  {Van~Assche}}, \bibinfo {author} {\bibfnamefont {J.}~\bibnamefont {Wenger}},
  \bibinfo {author} {\bibfnamefont {R.}~\bibnamefont {Brouri}}, \bibinfo
  {author} {\bibfnamefont {N.~J.}\ \bibnamefont {Cerf}},\ and\ \bibinfo
  {author} {\bibfnamefont {P.}~\bibnamefont {Grangier}},\ }\bibfield  {title}
  {\bibinfo {title} {Quantum key distribution using gaussian-modulated coherent
  states},\ }\href {https://doi.org/10.1038/nature01289} {\bibfield  {journal}
  {\bibinfo  {journal} {Nature}\ }\textbf {\bibinfo {volume} {421}},\ \bibinfo
  {pages} {238} (\bibinfo {year} {2003}{\natexlab{b}})}\BibitemShut {NoStop}%
\bibitem [{\citenamefont {Leverrier}\ \emph {et~al.}(2010)\citenamefont
  {Leverrier}, \citenamefont {Grosshans},\ and\ \citenamefont
  {Grangier}}]{Leverrier2010}%
  \BibitemOpen
  \bibfield  {author} {\bibinfo {author} {\bibfnamefont {A.}~\bibnamefont
  {Leverrier}}, \bibinfo {author} {\bibfnamefont {F.}~\bibnamefont
  {Grosshans}},\ and\ \bibinfo {author} {\bibfnamefont {P.}~\bibnamefont
  {Grangier}},\ }\bibfield  {title} {\bibinfo {title} {{Finite-size analysis of
  a continuous-variable quantum key distribution}},\ }\href
  {https://doi.org/10.1103/PhysRevA.81.062343} {\bibfield  {journal} {\bibinfo
  {journal} {Phys. Rev. A}\ }\textbf {\bibinfo {volume} {81}},\ \bibinfo
  {pages} {062343} (\bibinfo {year} {2010})}\BibitemShut {NoStop}%
\bibitem [{\citenamefont {Pirandola}\ \emph {et~al.}(2008)\citenamefont
  {Pirandola}, \citenamefont {Braunstein},\ and\ \citenamefont
  {Lloyd}}]{Pirandola2008}%
  \BibitemOpen
  \bibfield  {author} {\bibinfo {author} {\bibfnamefont {S.}~\bibnamefont
  {Pirandola}}, \bibinfo {author} {\bibfnamefont {S.~L.}\ \bibnamefont
  {Braunstein}},\ and\ \bibinfo {author} {\bibfnamefont {S.}~\bibnamefont
  {Lloyd}},\ }\bibfield  {title} {\bibinfo {title} {{Characterization of
  collective gaussian attacks and security of coherent-state quantum
  cryptography}},\ }\href {https://doi.org/10.1103/PhysRevLett.101.200504}
  {\bibfield  {journal} {\bibinfo  {journal} {Phys. Rev. Lett}\ }\textbf
  {\bibinfo {volume} {101}},\ \bibinfo {pages} {1} (\bibinfo {year}
  {2008})}\BibitemShut {NoStop}%
\bibitem [{\citenamefont {Grosshans}\ \emph
  {et~al.}(2003{\natexlab{c}})\citenamefont {Grosshans}, \citenamefont {Cerf},
  \citenamefont {Wenger}, \citenamefont {Tualle-Brouri},\ and\ \citenamefont
  {Grangier}}]{Grosshans2003_2}%
  \BibitemOpen
  \bibfield  {author} {\bibinfo {author} {\bibfnamefont {F.}~\bibnamefont
  {Grosshans}}, \bibinfo {author} {\bibfnamefont {N.~J.}\ \bibnamefont {Cerf}},
  \bibinfo {author} {\bibfnamefont {J.}~\bibnamefont {Wenger}}, \bibinfo
  {author} {\bibfnamefont {R.}~\bibnamefont {Tualle-Brouri}},\ and\ \bibinfo
  {author} {\bibfnamefont {P.}~\bibnamefont {Grangier}},\ }\bibfield  {title}
  {\bibinfo {title} {Virtual entanglement and reconciliation protocols for
  quantum cryptography with continuous variables},\ }\href
  {https://www.rintonpress.com/journals/doi/QIC3.s-6.html} {\bibfield
  {journal} {\bibinfo  {journal} {Quantum Info. Comput.}\ }\textbf {\bibinfo
  {volume} {3}},\ \bibinfo {pages} {535–552} (\bibinfo {year}
  {2003}{\natexlab{c}})}\BibitemShut {NoStop}%
\bibitem [{\citenamefont {Fedorov}\ \emph {et~al.}(2018)\citenamefont
  {Fedorov}, \citenamefont {Pogorzalek}, \citenamefont {{Las Heras}},
  \citenamefont {Sanz}, \citenamefont {Yard}, \citenamefont {Eder},
  \citenamefont {Fischer}, \citenamefont {Goetz}, \citenamefont {Xie},
  \citenamefont {Inomata}, \citenamefont {Nakamura}, \citenamefont {{Di
  Candia}}, \citenamefont {Solano}, \citenamefont {Marx}, \citenamefont
  {Deppe},\ and\ \citenamefont {Gross}}]{Fedorov2018}%
  \BibitemOpen
  \bibfield  {author} {\bibinfo {author} {\bibfnamefont {K.~G.}\ \bibnamefont
  {Fedorov}}, \bibinfo {author} {\bibfnamefont {S.}~\bibnamefont {Pogorzalek}},
  \bibinfo {author} {\bibfnamefont {U.}~\bibnamefont {{Las Heras}}}, \bibinfo
  {author} {\bibfnamefont {M.}~\bibnamefont {Sanz}}, \bibinfo {author}
  {\bibfnamefont {P.}~\bibnamefont {Yard}}, \bibinfo {author} {\bibfnamefont
  {P.}~\bibnamefont {Eder}}, \bibinfo {author} {\bibfnamefont {M.}~\bibnamefont
  {Fischer}}, \bibinfo {author} {\bibfnamefont {J.}~\bibnamefont {Goetz}},
  \bibinfo {author} {\bibfnamefont {E.}~\bibnamefont {Xie}}, \bibinfo {author}
  {\bibfnamefont {K.}~\bibnamefont {Inomata}}, \bibinfo {author} {\bibfnamefont
  {Y.}~\bibnamefont {Nakamura}}, \bibinfo {author} {\bibfnamefont
  {R.}~\bibnamefont {{Di Candia}}}, \bibinfo {author} {\bibfnamefont
  {E.}~\bibnamefont {Solano}}, \bibinfo {author} {\bibfnamefont
  {A.}~\bibnamefont {Marx}}, \bibinfo {author} {\bibfnamefont {F.}~\bibnamefont
  {Deppe}},\ and\ \bibinfo {author} {\bibfnamefont {R.}~\bibnamefont {Gross}},\
  }\bibfield  {title} {\bibinfo {title} {{Finite-time quantum entanglement in
  propagating squeezed microwaves}},\ }\href
  {https://doi.org/10.1038/s41598-018-24742-z} {\bibfield  {journal} {\bibinfo
  {journal} {Sci. Rep.}\ }\textbf {\bibinfo {volume} {8}},\ \bibinfo {pages}
  {1} (\bibinfo {year} {2018})}\BibitemShut {NoStop}%
\bibitem [{\citenamefont {Fedorov}\ \emph {et~al.}(2016)\citenamefont
  {Fedorov}, \citenamefont {Zhong}, \citenamefont {Pogorzalek}, \citenamefont
  {Eder}, \citenamefont {Fischer}, \citenamefont {Goetz}, \citenamefont {Xie},
  \citenamefont {Wulschner}, \citenamefont {Inomata}, \citenamefont {Yamamoto},
  \citenamefont {Nakamura}, \citenamefont {{Di Candia}}, \citenamefont {{Las
  Heras}}, \citenamefont {Sanz}, \citenamefont {Solano}, \citenamefont
  {Menzel}, \citenamefont {Deppe}, \citenamefont {Marx},\ and\ \citenamefont
  {Gross}}]{Fedorov2016}%
  \BibitemOpen
  \bibfield  {author} {\bibinfo {author} {\bibfnamefont {K.~G.}\ \bibnamefont
  {Fedorov}}, \bibinfo {author} {\bibfnamefont {L.}~\bibnamefont {Zhong}},
  \bibinfo {author} {\bibfnamefont {S.}~\bibnamefont {Pogorzalek}}, \bibinfo
  {author} {\bibfnamefont {P.}~\bibnamefont {Eder}}, \bibinfo {author}
  {\bibfnamefont {M.}~\bibnamefont {Fischer}}, \bibinfo {author} {\bibfnamefont
  {J.}~\bibnamefont {Goetz}}, \bibinfo {author} {\bibfnamefont
  {E.}~\bibnamefont {Xie}}, \bibinfo {author} {\bibfnamefont {F.}~\bibnamefont
  {Wulschner}}, \bibinfo {author} {\bibfnamefont {K.}~\bibnamefont {Inomata}},
  \bibinfo {author} {\bibfnamefont {T.}~\bibnamefont {Yamamoto}}, \bibinfo
  {author} {\bibfnamefont {Y.}~\bibnamefont {Nakamura}}, \bibinfo {author}
  {\bibfnamefont {R.}~\bibnamefont {{Di Candia}}}, \bibinfo {author}
  {\bibfnamefont {U.}~\bibnamefont {{Las Heras}}}, \bibinfo {author}
  {\bibfnamefont {M.}~\bibnamefont {Sanz}}, \bibinfo {author} {\bibfnamefont
  {E.}~\bibnamefont {Solano}}, \bibinfo {author} {\bibfnamefont {E.~P.}\
  \bibnamefont {Menzel}}, \bibinfo {author} {\bibfnamefont {F.}~\bibnamefont
  {Deppe}}, \bibinfo {author} {\bibfnamefont {A.}~\bibnamefont {Marx}},\ and\
  \bibinfo {author} {\bibfnamefont {R.}~\bibnamefont {Gross}},\ }\bibfield
  {title} {\bibinfo {title} {{Displacement of Propagating Squeezed Microwave
  States}},\ }\href {https://doi.org/10.1103/PhysRevLett.117.020502} {\bibfield
   {journal} {\bibinfo  {journal} {Phys. Rev. Lett}\ }\textbf {\bibinfo
  {volume} {117}},\ \bibinfo {pages} {020502} (\bibinfo {year}
  {2016})}\BibitemShut {NoStop}%
\bibitem [{\citenamefont {Yamamoto}\ \emph {et~al.}(2008)\citenamefont
  {Yamamoto}, \citenamefont {Inomata}, \citenamefont {Watanabe}, \citenamefont
  {Matsuba}, \citenamefont {Miyazaki}, \citenamefont {Oliver}, \citenamefont
  {Nakamura},\ and\ \citenamefont {Tsai}}]{Yamamoto2008}%
  \BibitemOpen
  \bibfield  {author} {\bibinfo {author} {\bibfnamefont {T.}~\bibnamefont
  {Yamamoto}}, \bibinfo {author} {\bibfnamefont {K.}~\bibnamefont {Inomata}},
  \bibinfo {author} {\bibfnamefont {M.}~\bibnamefont {Watanabe}}, \bibinfo
  {author} {\bibfnamefont {K.}~\bibnamefont {Matsuba}}, \bibinfo {author}
  {\bibfnamefont {T.}~\bibnamefont {Miyazaki}}, \bibinfo {author}
  {\bibfnamefont {W.~D.}\ \bibnamefont {Oliver}}, \bibinfo {author}
  {\bibfnamefont {Y.}~\bibnamefont {Nakamura}},\ and\ \bibinfo {author}
  {\bibfnamefont {J.~S.}\ \bibnamefont {Tsai}},\ }\bibfield  {title} {\bibinfo
  {title} {{Flux-driven Josephson parametric amplifier}},\ }\href
  {https://doi.org/10.1063/1.2964182} {\bibfield  {journal} {\bibinfo
  {journal} {Appl. Phys. Lett.}\ }\textbf {\bibinfo {volume} {93}},\ \bibinfo
  {pages} {042510} (\bibinfo {year} {2008})}\BibitemShut {NoStop}%
\bibitem [{\citenamefont {Yurke}\ \emph {et~al.}(1989)\citenamefont {Yurke},
  \citenamefont {Corruccini}, \citenamefont {Kaminsky}, \citenamefont {Rupp},
  \citenamefont {Smith}, \citenamefont {Silver}, \citenamefont {Simon},\ and\
  \citenamefont {Whittaker}}]{Yurke1989}%
  \BibitemOpen
  \bibfield  {author} {\bibinfo {author} {\bibfnamefont {B.}~\bibnamefont
  {Yurke}}, \bibinfo {author} {\bibfnamefont {L.~R.}\ \bibnamefont
  {Corruccini}}, \bibinfo {author} {\bibfnamefont {P.~G.}\ \bibnamefont
  {Kaminsky}}, \bibinfo {author} {\bibfnamefont {L.~W.}\ \bibnamefont {Rupp}},
  \bibinfo {author} {\bibfnamefont {A.~D.}\ \bibnamefont {Smith}}, \bibinfo
  {author} {\bibfnamefont {A.~H.}\ \bibnamefont {Silver}}, \bibinfo {author}
  {\bibfnamefont {R.~W.}\ \bibnamefont {Simon}},\ and\ \bibinfo {author}
  {\bibfnamefont {E.~A.}\ \bibnamefont {Whittaker}},\ }\bibfield  {title}
  {\bibinfo {title} {{Observation of parametric amplification and
  deamplification in a Josephson parametric amplifier}},\ }\href
  {https://doi.org/10.1103/PhysRevA.39.2519} {\bibfield  {journal} {\bibinfo
  {journal} {Phys. Rev. A}\ }\textbf {\bibinfo {volume} {39}},\ \bibinfo
  {pages} {2519} (\bibinfo {year} {1989})}\BibitemShut {NoStop}%
\bibitem [{\citenamefont {Caves}(1982)}]{Caves1982}%
  \BibitemOpen
  \bibfield  {author} {\bibinfo {author} {\bibfnamefont {C.~M.}\ \bibnamefont
  {Caves}},\ }\bibfield  {title} {\bibinfo {title} {{Quantum limits on noise in
  linear amplifiers}},\ }\href {https://doi.org/10.1103/PhysRevD.26.1817}
  {\bibfield  {journal} {\bibinfo  {journal} {Phys. Rev. D}\ }\textbf {\bibinfo
  {volume} {26}},\ \bibinfo {pages} {1817} (\bibinfo {year}
  {1982})}\BibitemShut {NoStop}%
\bibitem [{\citenamefont {Renger}\ \emph {et~al.}(2021)\citenamefont {Renger},
  \citenamefont {Pogorzalek}, \citenamefont {Chen}, \citenamefont {Nojiri},
  \citenamefont {Inomata}, \citenamefont {Nakamura}, \citenamefont {Partanen},
  \citenamefont {Marx}, \citenamefont {Gross}, \citenamefont {Deppe},\ and\
  \citenamefont {Fedorov}}]{Renger2021}%
  \BibitemOpen
  \bibfield  {author} {\bibinfo {author} {\bibfnamefont {M.}~\bibnamefont
  {Renger}}, \bibinfo {author} {\bibfnamefont {S.}~\bibnamefont {Pogorzalek}},
  \bibinfo {author} {\bibfnamefont {Q.}~\bibnamefont {Chen}}, \bibinfo {author}
  {\bibfnamefont {Y.}~\bibnamefont {Nojiri}}, \bibinfo {author} {\bibfnamefont
  {K.}~\bibnamefont {Inomata}}, \bibinfo {author} {\bibfnamefont
  {Y.}~\bibnamefont {Nakamura}}, \bibinfo {author} {\bibfnamefont
  {M.}~\bibnamefont {Partanen}}, \bibinfo {author} {\bibfnamefont
  {A.}~\bibnamefont {Marx}}, \bibinfo {author} {\bibfnamefont {R.}~\bibnamefont
  {Gross}}, \bibinfo {author} {\bibfnamefont {F.}~\bibnamefont {Deppe}},\ and\
  \bibinfo {author} {\bibfnamefont {K.~G.}\ \bibnamefont {Fedorov}},\
  }\bibfield  {title} {\bibinfo {title} {{Beyond the standard quantum limit for
  parametric amplification of broadband signals}},\ }\href
  {https://doi.org/10.1038/s41534-021-00495-y} {\bibfield  {journal} {\bibinfo
  {journal} {Npj Quantum Inf.}\ }\textbf {\bibinfo {volume} {7}},\ \bibinfo
  {pages} {160} (\bibinfo {year} {2021})}\BibitemShut {NoStop}%
\bibitem [{\citenamefont {Fedorov}\ \emph {et~al.}(2021)\citenamefont
  {Fedorov}, \citenamefont {Renger}, \citenamefont {Pogorzalek}, \citenamefont
  {Candia}, \citenamefont {Chen}, \citenamefont {Nojiri}, \citenamefont
  {Inomata}, \citenamefont {Nakamura}, \citenamefont {Partanen}, \citenamefont
  {Marx}, \citenamefont {Gross},\ and\ \citenamefont {Deppe}}]{Fedorov21}%
  \BibitemOpen
  \bibfield  {author} {\bibinfo {author} {\bibfnamefont {K.~G.}\ \bibnamefont
  {Fedorov}}, \bibinfo {author} {\bibfnamefont {M.}~\bibnamefont {Renger}},
  \bibinfo {author} {\bibfnamefont {S.}~\bibnamefont {Pogorzalek}}, \bibinfo
  {author} {\bibfnamefont {R.~D.}\ \bibnamefont {Candia}}, \bibinfo {author}
  {\bibfnamefont {Q.}~\bibnamefont {Chen}}, \bibinfo {author} {\bibfnamefont
  {Y.}~\bibnamefont {Nojiri}}, \bibinfo {author} {\bibfnamefont
  {K.}~\bibnamefont {Inomata}}, \bibinfo {author} {\bibfnamefont
  {Y.}~\bibnamefont {Nakamura}}, \bibinfo {author} {\bibfnamefont
  {M.}~\bibnamefont {Partanen}}, \bibinfo {author} {\bibfnamefont
  {A.}~\bibnamefont {Marx}}, \bibinfo {author} {\bibfnamefont {R.}~\bibnamefont
  {Gross}},\ and\ \bibinfo {author} {\bibfnamefont {F.}~\bibnamefont {Deppe}},\
  }\bibfield  {title} {\bibinfo {title} {Experimental quantum teleportation of
  propagating microwaves},\ }\href {https://doi.org/10.1126/sciadv.abk0891}
  {\bibfield  {journal} {\bibinfo  {journal} {Science Advances}\ }\textbf
  {\bibinfo {volume} {7}},\ \bibinfo {pages} {eabk0891} (\bibinfo {year}
  {2021})}\BibitemShut {NoStop}%
\bibitem [{\citenamefont {Boutin}\ \emph {et~al.}(2017)\citenamefont {Boutin},
  \citenamefont {Toyli}, \citenamefont {Venkatramani}, \citenamefont {Eddins},
  \citenamefont {Siddiqi},\ and\ \citenamefont {Blais}}]{Boutin2017}%
  \BibitemOpen
  \bibfield  {author} {\bibinfo {author} {\bibfnamefont {S.}~\bibnamefont
  {Boutin}}, \bibinfo {author} {\bibfnamefont {D.~M.}\ \bibnamefont {Toyli}},
  \bibinfo {author} {\bibfnamefont {A.~V.}\ \bibnamefont {Venkatramani}},
  \bibinfo {author} {\bibfnamefont {A.~W.}\ \bibnamefont {Eddins}}, \bibinfo
  {author} {\bibfnamefont {I.}~\bibnamefont {Siddiqi}},\ and\ \bibinfo {author}
  {\bibfnamefont {A.}~\bibnamefont {Blais}},\ }\bibfield  {title} {\bibinfo
  {title} {{Effect of Higher-Order Nonlinearities on Amplification and
  Squeezing in Josephson Parametric Amplifiers}},\ }\href
  {https://doi.org/10.1103/PhysRevApplied.8.054030} {\bibfield  {journal}
  {\bibinfo  {journal} {Phys. Rev. Appl.}\ }\textbf {\bibinfo {volume} {8}},\
  \bibinfo {pages} {054030} (\bibinfo {year} {2017})}\BibitemShut {NoStop}%
\bibitem [{\citenamefont {Pozar}(2011)}]{Pozar2011}%
  \BibitemOpen
  \bibfield  {author} {\bibinfo {author} {\bibfnamefont {D.~M.}\ \bibnamefont
  {Pozar}},\ }\href
  {https://www.wiley.com/en-us/Microwave+Engineering,+4th+Edition-p-9780470631553}
  {\emph {\bibinfo {title} {{Microwave engineering; 4th ed.}}}}\ (\bibinfo
  {publisher} {Wiley},\ \bibinfo {address} {Hoboken},\ \bibinfo {year}
  {2011})\BibitemShut {NoStop}%
\bibitem [{\citenamefont {Gonzalez-Raya}\ and\ \citenamefont
  {Sanz}(2020)}]{gonzalezraya2020coplanar}%
  \BibitemOpen
  \bibfield  {author} {\bibinfo {author} {\bibfnamefont {T.}~\bibnamefont
  {Gonzalez-Raya}}\ and\ \bibinfo {author} {\bibfnamefont {M.}~\bibnamefont
  {Sanz}},\ }\bibfield  {title} {\bibinfo {title} {Coplanar antenna design for
  microwave entangled signals propagating in open air},\ }\href
  {https://arxiv.org/abs/2009.03021} {\  (\bibinfo {year} {2020})},\ \Eprint
  {https://arxiv.org/abs/2009.03021} {arXiv:2009.03021 [quant-ph]} \BibitemShut
  {NoStop}%
\bibitem [{\citenamefont {Pirandola}(2021)}]{Pirandola2021}%
  \BibitemOpen
  \bibfield  {author} {\bibinfo {author} {\bibfnamefont {S.}~\bibnamefont
  {Pirandola}},\ }\bibfield  {title} {\bibinfo {title} {{Composable security
  for continuous variable quantum key distribution: Trust levels and practical
  key rates in wired and wireless networks}},\ }\href
  {https://doi.org/10.1103/PhysRevResearch.3.043014} {\bibfield  {journal}
  {\bibinfo  {journal} {Physical Review Research}\ }\textbf {\bibinfo {volume}
  {3}},\ \bibinfo {pages} {043014} (\bibinfo {year} {2021})},\ \Eprint
  {https://arxiv.org/abs/2203.00706} {2203.00706} \BibitemShut {NoStop}%
\bibitem [{\citenamefont {Castellanos-Beltran}\ \emph
  {et~al.}(2008)\citenamefont {Castellanos-Beltran}, \citenamefont {Irwin},
  \citenamefont {Hilton}, \citenamefont {Vale},\ and\ \citenamefont
  {Lehnert}}]{Castellanos-Beltran2008}%
  \BibitemOpen
  \bibfield  {author} {\bibinfo {author} {\bibfnamefont {M.~A.}\ \bibnamefont
  {Castellanos-Beltran}}, \bibinfo {author} {\bibfnamefont {K.~D.}\
  \bibnamefont {Irwin}}, \bibinfo {author} {\bibfnamefont {G.~C.}\ \bibnamefont
  {Hilton}}, \bibinfo {author} {\bibfnamefont {L.~R.}\ \bibnamefont {Vale}},\
  and\ \bibinfo {author} {\bibfnamefont {K.~W.}\ \bibnamefont {Lehnert}},\
  }\bibfield  {title} {\bibinfo {title} {{Amplification and squeezing of
  quantum noise with a tunable Josephson metamaterial}},\ }\href
  {https://doi.org/10.1038/nphys1090} {\bibfield  {journal} {\bibinfo
  {journal} {Nat. Phys}\ }\textbf {\bibinfo {volume} {4}},\ \bibinfo {pages}
  {929} (\bibinfo {year} {2008})}\BibitemShut {NoStop}%
\bibitem [{\citenamefont {Eichler}\ \emph {et~al.}(2012)\citenamefont
  {Eichler}, \citenamefont {Bozyigit},\ and\ \citenamefont
  {Wallraff}}]{Eichler2012}%
  \BibitemOpen
  \bibfield  {author} {\bibinfo {author} {\bibfnamefont {C.}~\bibnamefont
  {Eichler}}, \bibinfo {author} {\bibfnamefont {D.}~\bibnamefont {Bozyigit}},\
  and\ \bibinfo {author} {\bibfnamefont {A.}~\bibnamefont {Wallraff}},\
  }\bibfield  {title} {\bibinfo {title} {{Characterizing quantum microwave
  radiation and its entanglement with superconducting qubits using linear
  detectors}},\ }\href {https://doi.org/10.1103/PhysRevA.86.032106} {\bibfield
  {journal} {\bibinfo  {journal} {Phys. Rev. A}\ }\textbf {\bibinfo {volume}
  {86}},\ \bibinfo {pages} {032106} (\bibinfo {year} {2012})}\BibitemShut
  {NoStop}%
\bibitem [{\citenamefont {{Ho}}\ \emph {et~al.}(2004)\citenamefont {{Ho}},
  \citenamefont {{Wang}}, \citenamefont {{Angkasa}},\ and\ \citenamefont
  {{Gritton}}}]{Ho2004}%
  \BibitemOpen
  \bibfield  {author} {\bibinfo {author} {\bibfnamefont {C.~M.}\ \bibnamefont
  {{Ho}}}, \bibinfo {author} {\bibfnamefont {C.}~\bibnamefont {{Wang}}},
  \bibinfo {author} {\bibfnamefont {K.}~\bibnamefont {{Angkasa}}},\ and\
  \bibinfo {author} {\bibfnamefont {K.}~\bibnamefont {{Gritton}}},\ }\bibfield
  {title} {\bibinfo {title} {Estimation of microwave power margin losses due to
  earth’s atmosphere and weather in the frequency range of 3–30 ghz},\
  }\href {https://descanso.jpl.nasa.gov/propagation/Ka_Band/JPL_D27879.pdf}
  {\bibfield  {journal} {\bibinfo  {journal} {Jet Propulsion Laboratory}\ }
  (\bibinfo {year} {2004})}\BibitemShut {NoStop}%
\bibitem [{\citenamefont {Holevo}(1973)}]{Holevo1973}%
  \BibitemOpen
  \bibfield  {author} {\bibinfo {author} {\bibfnamefont {A.~S.}\ \bibnamefont
  {Holevo}},\ }\bibfield  {title} {\bibinfo {title} {{Bounds for the quantity
  of information transmitted by a quantum communication channel}},\ }\href
  {http://www.mathnet.ru/php/archive.phtml?wshow=paper&jrnid=ppi&paperid=903&option_lang=eng}
  {\bibfield  {journal} {\bibinfo  {journal} {Probl. Inf. Transm.}\ }\textbf
  {\bibinfo {volume} {9}},\ \bibinfo {pages} {177} (\bibinfo {year}
  {1973})}\BibitemShut {NoStop}%
\bibitem [{\citenamefont {Lupo}\ \emph {et~al.}(2018)\citenamefont {Lupo},
  \citenamefont {Ottaviani}, \citenamefont {Papanastasiou},\ and\ \citenamefont
  {Pirandola}}]{Pirandola2018composable}%
  \BibitemOpen
  \bibfield  {author} {\bibinfo {author} {\bibfnamefont {C.}~\bibnamefont
  {Lupo}}, \bibinfo {author} {\bibfnamefont {C.}~\bibnamefont {Ottaviani}},
  \bibinfo {author} {\bibfnamefont {P.}~\bibnamefont {Papanastasiou}},\ and\
  \bibinfo {author} {\bibfnamefont {S.}~\bibnamefont {Pirandola}},\ }\bibfield
  {title} {\bibinfo {title} {Continuous-variable measurement-device-independent
  quantum key distribution: Composable security against coherent attacks},\
  }\href {https://doi.org/10.1103/PhysRevA.97.052327} {\bibfield  {journal}
  {\bibinfo  {journal} {Phys. Rev. A}\ }\textbf {\bibinfo {volume} {97}},\
  \bibinfo {pages} {052327} (\bibinfo {year} {2018})}\BibitemShut {NoStop}%
\bibitem [{\citenamefont {Grosshans}\ and\ \citenamefont
  {Grangier}(2002)}]{grosshans2002}%
  \BibitemOpen
  \bibfield  {author} {\bibinfo {author} {\bibfnamefont {F.}~\bibnamefont
  {Grosshans}}\ and\ \bibinfo {author} {\bibfnamefont {P.}~\bibnamefont
  {Grangier}},\ }\bibfield  {title} {\bibinfo {title} {Continuous variable
  quantum cryptography using coherent states},\ }\href
  {https://doi.org/10.1103/PhysRevLett.88.057902} {\bibfield  {journal}
  {\bibinfo  {journal} {Phys. Rev. Lett}\ }\textbf {\bibinfo {volume} {88}},\
  \bibinfo {pages} {057902} (\bibinfo {year} {2002})}\BibitemShut {NoStop}%
\bibitem [{\citenamefont {Weedbrook}\ \emph {et~al.}(2012)\citenamefont
  {Weedbrook}, \citenamefont {Pirandola},\ and\ \citenamefont
  {Ralph}}]{Weedbrook2012}%
  \BibitemOpen
  \bibfield  {author} {\bibinfo {author} {\bibfnamefont {C.}~\bibnamefont
  {Weedbrook}}, \bibinfo {author} {\bibfnamefont {S.}~\bibnamefont
  {Pirandola}},\ and\ \bibinfo {author} {\bibfnamefont {T.~C.}\ \bibnamefont
  {Ralph}},\ }\bibfield  {title} {\bibinfo {title} {{Continuous-variable
  quantum key distribution using thermal states}},\ }\href
  {https://doi.org/10.1103/PhysRevA.86.022318} {\bibfield  {journal} {\bibinfo
  {journal} {Phys. Rev. A}\ }\textbf {\bibinfo {volume} {86}},\ \bibinfo
  {pages} {022318} (\bibinfo {year} {2012})}\BibitemShut {NoStop}%
\bibitem [{\citenamefont {Pirandola}\ \emph {et~al.}(2017)\citenamefont
  {Pirandola}, \citenamefont {Laurenza}, \citenamefont {Ottaviani},\ and\
  \citenamefont {Banchi}}]{Pirandola2017}%
  \BibitemOpen
  \bibfield  {author} {\bibinfo {author} {\bibfnamefont {S.}~\bibnamefont
  {Pirandola}}, \bibinfo {author} {\bibfnamefont {R.}~\bibnamefont {Laurenza}},
  \bibinfo {author} {\bibfnamefont {C.}~\bibnamefont {Ottaviani}},\ and\
  \bibinfo {author} {\bibfnamefont {L.}~\bibnamefont {Banchi}},\ }\bibfield
  {title} {\bibinfo {title} {{Fundamental limits of repeaterless quantum
  communications}},\ }\href {https://doi.org/10.1038/ncomms15043} {\bibfield
  {journal} {\bibinfo  {journal} {Nature Communications}\ }\textbf {\bibinfo
  {volume} {8}},\ \bibinfo {pages} {15043} (\bibinfo {year}
  {2017})}\BibitemShut {NoStop}%
\bibitem [{\citenamefont {Nyquist}(1928)}]{nyquist1928}%
  \BibitemOpen
  \bibfield  {author} {\bibinfo {author} {\bibfnamefont {H.}~\bibnamefont
  {Nyquist}},\ }\bibfield  {title} {\bibinfo {title} {Certain topics in
  telegraph transmission theory},\ }\href
  {https://doi.org/10.1109/T-AIEE.1928.5055024} {\bibfield  {journal} {\bibinfo
   {journal} {Trans. Am. Inst. Electr. Eng.}\ }\textbf {\bibinfo {volume}
  {47}},\ \bibinfo {pages} {617} (\bibinfo {year} {1928})}\BibitemShut
  {NoStop}%
\bibitem [{\citenamefont {Ast}\ \emph {et~al.}(2013)\citenamefont {Ast},
  \citenamefont {Mehmet},\ and\ \citenamefont {Schnabel}}]{Ast2013}%
  \BibitemOpen
  \bibfield  {author} {\bibinfo {author} {\bibfnamefont {S.}~\bibnamefont
  {Ast}}, \bibinfo {author} {\bibfnamefont {M.}~\bibnamefont {Mehmet}},\ and\
  \bibinfo {author} {\bibfnamefont {R.}~\bibnamefont {Schnabel}},\ }\bibfield
  {title} {\bibinfo {title} {{High-bandwidth squeezed light at 1550 nm from a
  compact monolithic PPKTP cavity}},\ }\href
  {https://doi.org/10.1364/OE.21.013572} {\bibfield  {journal} {\bibinfo
  {journal} {Opt. Express}\ }\textbf {\bibinfo {volume} {21}},\ \bibinfo
  {pages} {13572} (\bibinfo {year} {2013})}\BibitemShut {NoStop}%
\bibitem [{\citenamefont {Garc{\'{i}}a-Patr{\'{o}}n}\ and\ \citenamefont
  {Cerf}(2009)}]{Garcia-Patron2009}%
  \BibitemOpen
  \bibfield  {author} {\bibinfo {author} {\bibfnamefont {R.}~\bibnamefont
  {Garc{\'{i}}a-Patr{\'{o}}n}}\ and\ \bibinfo {author} {\bibfnamefont {N.~J.}\
  \bibnamefont {Cerf}},\ }\bibfield  {title} {\bibinfo {title}
  {{Continuous-Variable Quantum Key Distribution Protocols Over Noisy
  Channels}},\ }\href {https://doi.org/10.1103/PhysRevLett.102.130501}
  {\bibfield  {journal} {\bibinfo  {journal} {Phys. Rev. Lett}\ }\textbf
  {\bibinfo {volume} {102}},\ \bibinfo {pages} {130501} (\bibinfo {year}
  {2009})}\BibitemShut {NoStop}%
\bibitem [{\citenamefont {Macklin}\ \emph {et~al.}(2015)\citenamefont
  {Macklin}, \citenamefont {O'Brien}, \citenamefont {Hover}, \citenamefont
  {Schwartz}, \citenamefont {Bolkhovsky}, \citenamefont {Zhang}, \citenamefont
  {Oliver},\ and\ \citenamefont {Siddiqi}}]{Macklin2015}%
  \BibitemOpen
  \bibfield  {author} {\bibinfo {author} {\bibfnamefont {C.}~\bibnamefont
  {Macklin}}, \bibinfo {author} {\bibfnamefont {K.}~\bibnamefont {O'Brien}},
  \bibinfo {author} {\bibfnamefont {D.}~\bibnamefont {Hover}}, \bibinfo
  {author} {\bibfnamefont {M.~E.}\ \bibnamefont {Schwartz}}, \bibinfo {author}
  {\bibfnamefont {V.}~\bibnamefont {Bolkhovsky}}, \bibinfo {author}
  {\bibfnamefont {X.}~\bibnamefont {Zhang}}, \bibinfo {author} {\bibfnamefont
  {W.~D.}\ \bibnamefont {Oliver}},\ and\ \bibinfo {author} {\bibfnamefont
  {I.}~\bibnamefont {Siddiqi}},\ }\bibfield  {title} {\bibinfo {title} {{A
  near-quantum-limited Josephson traveling-wave parametric amplifier}},\ }\href
  {https://doi.org/10.1126/science.aaa8525} {\bibfield  {journal} {\bibinfo
  {journal} {Science}\ }\textbf {\bibinfo {volume} {350}},\ \bibinfo {pages}
  {307} (\bibinfo {year} {2015})}\BibitemShut {NoStop}%
\bibitem [{\citenamefont {Perelshtein}\ \emph {et~al.}(2021)\citenamefont
  {Perelshtein}, \citenamefont {Petrovnin}, \citenamefont {Vesterinen},
  \citenamefont {Raja}, \citenamefont {Lilja}, \citenamefont {Will},
  \citenamefont {Savin}, \citenamefont {Simbierowicz}, \citenamefont
  {Jabdaraghi}, \citenamefont {Lehtinen}, \citenamefont {Gr{\"{o}}nberg},
  \citenamefont {Hassel}, \citenamefont {Prunnila}, \citenamefont {Govenius},
  \citenamefont {Paraoanu},\ and\ \citenamefont {Hakonen}}]{Perelshtein2021}%
  \BibitemOpen
  \bibfield  {author} {\bibinfo {author} {\bibfnamefont {M.}~\bibnamefont
  {Perelshtein}}, \bibinfo {author} {\bibfnamefont {K.}~\bibnamefont
  {Petrovnin}}, \bibinfo {author} {\bibfnamefont {V.}~\bibnamefont
  {Vesterinen}}, \bibinfo {author} {\bibfnamefont {S.~H.}\ \bibnamefont
  {Raja}}, \bibinfo {author} {\bibfnamefont {I.}~\bibnamefont {Lilja}},
  \bibinfo {author} {\bibfnamefont {M.}~\bibnamefont {Will}}, \bibinfo {author}
  {\bibfnamefont {A.}~\bibnamefont {Savin}}, \bibinfo {author} {\bibfnamefont
  {S.}~\bibnamefont {Simbierowicz}}, \bibinfo {author} {\bibfnamefont
  {R.}~\bibnamefont {Jabdaraghi}}, \bibinfo {author} {\bibfnamefont
  {J.}~\bibnamefont {Lehtinen}}, \bibinfo {author} {\bibfnamefont
  {L.}~\bibnamefont {Gr{\"{o}}nberg}}, \bibinfo {author} {\bibfnamefont
  {J.}~\bibnamefont {Hassel}}, \bibinfo {author} {\bibfnamefont
  {M.}~\bibnamefont {Prunnila}}, \bibinfo {author} {\bibfnamefont
  {J.}~\bibnamefont {Govenius}}, \bibinfo {author} {\bibfnamefont
  {S.}~\bibnamefont {Paraoanu}},\ and\ \bibinfo {author} {\bibfnamefont
  {P.}~\bibnamefont {Hakonen}},\ }\bibfield  {title} {\bibinfo {title}
  {{Broadband continuous variable entanglement generation using Kerr-free
  Josephson metamaterial}},\ }\href {http://arxiv.org/abs/2111.06145} {\
  (\bibinfo {year} {2021})},\ \Eprint {https://arxiv.org/abs/2111.06145}
  {arXiv:2111.06145 [quant-ph]} \BibitemShut {NoStop}%
\bibitem [{\citenamefont {Schneider}\ \emph {et~al.}(2020)\citenamefont
  {Schneider}, \citenamefont {Bengtsson}, \citenamefont {Svensson},
  \citenamefont {Aref}, \citenamefont {Johansson}, \citenamefont {Bylander},\
  and\ \citenamefont {Delsing}}]{Schneider2020}%
  \BibitemOpen
  \bibfield  {author} {\bibinfo {author} {\bibfnamefont {B.~H.}\ \bibnamefont
  {Schneider}}, \bibinfo {author} {\bibfnamefont {A.}~\bibnamefont
  {Bengtsson}}, \bibinfo {author} {\bibfnamefont {I.~M.}\ \bibnamefont
  {Svensson}}, \bibinfo {author} {\bibfnamefont {T.}~\bibnamefont {Aref}},
  \bibinfo {author} {\bibfnamefont {G.}~\bibnamefont {Johansson}}, \bibinfo
  {author} {\bibfnamefont {J.}~\bibnamefont {Bylander}},\ and\ \bibinfo
  {author} {\bibfnamefont {P.}~\bibnamefont {Delsing}},\ }\bibfield  {title}
  {\bibinfo {title} {{Observation of Broadband Entanglement in Microwave
  Radiation from a Single Time-Varying Boundary Condition}},\ }\href
  {https://doi.org/10.1103/PhysRevLett.124.140503} {\bibfield  {journal}
  {\bibinfo  {journal} {Phys. Rev. Lett}\ }\textbf {\bibinfo {volume} {124}},\
  \bibinfo {pages} {140503} (\bibinfo {year} {2020})}\BibitemShut {NoStop}%
\bibitem [{\citenamefont {Qiu}\ \emph {et~al.}(2022)\citenamefont {Qiu},
  \citenamefont {Grimsmo}, \citenamefont {Peng}, \citenamefont {Kannan},
  \citenamefont {Lienhard}, \citenamefont {Sung}, \citenamefont {Krantz},
  \citenamefont {Bolkhovsky}, \citenamefont {Calusine}, \citenamefont {Kim},
  \citenamefont {Melville}, \citenamefont {Niedzielski}, \citenamefont {Yoder},
  \citenamefont {Schwartz}, \citenamefont {Orlando}, \citenamefont {Siddiqi},
  \citenamefont {Gustavsson}, \citenamefont {O~Brien},\ and\ \citenamefont
  {Oliver}}]{Qiu2022}%
  \BibitemOpen
  \bibfield  {author} {\bibinfo {author} {\bibfnamefont {J.~Y.}\ \bibnamefont
  {Qiu}}, \bibinfo {author} {\bibfnamefont {A.}~\bibnamefont {Grimsmo}},
  \bibinfo {author} {\bibfnamefont {K.}~\bibnamefont {Peng}}, \bibinfo {author}
  {\bibfnamefont {B.}~\bibnamefont {Kannan}}, \bibinfo {author} {\bibfnamefont
  {B.}~\bibnamefont {Lienhard}}, \bibinfo {author} {\bibfnamefont
  {Y.}~\bibnamefont {Sung}}, \bibinfo {author} {\bibfnamefont {P.}~\bibnamefont
  {Krantz}}, \bibinfo {author} {\bibfnamefont {V.}~\bibnamefont {Bolkhovsky}},
  \bibinfo {author} {\bibfnamefont {G.}~\bibnamefont {Calusine}}, \bibinfo
  {author} {\bibfnamefont {D.}~\bibnamefont {Kim}}, \bibinfo {author}
  {\bibfnamefont {A.}~\bibnamefont {Melville}}, \bibinfo {author}
  {\bibfnamefont {B.~M.}\ \bibnamefont {Niedzielski}}, \bibinfo {author}
  {\bibfnamefont {J.}~\bibnamefont {Yoder}}, \bibinfo {author} {\bibfnamefont
  {M.~E.}\ \bibnamefont {Schwartz}}, \bibinfo {author} {\bibfnamefont {T.~P.}\
  \bibnamefont {Orlando}}, \bibinfo {author} {\bibfnamefont {I.}~\bibnamefont
  {Siddiqi}}, \bibinfo {author} {\bibfnamefont {S.}~\bibnamefont {Gustavsson}},
  \bibinfo {author} {\bibfnamefont {K.~P.}\ \bibnamefont {O~Brien}},\ and\
  \bibinfo {author} {\bibfnamefont {W.~D.}\ \bibnamefont {Oliver}},\ }\bibfield
   {title} {\bibinfo {title} {{Broadband Squeezed Microwaves and Amplification
  with a Josephson Traveling Wave Parametric Amplifier}},\ }\href
  {http://arxiv.org/abs/2201.11261} {\  (\bibinfo {year} {2022})},\ \Eprint
  {https://arxiv.org/abs/2201.11261} {arXiv:2201.11261} \BibitemShut {NoStop}%
\bibitem [{\citenamefont {Huang}\ \emph {et~al.}(2015)\citenamefont {Huang},
  \citenamefont {Lin}, \citenamefont {Wang}, \citenamefont {Liu}, \citenamefont
  {Fang}, \citenamefont {Peng}, \citenamefont {Huang},\ and\ \citenamefont
  {Zeng}}]{Huang2015}%
  \BibitemOpen
  \bibfield  {author} {\bibinfo {author} {\bibfnamefont {D.}~\bibnamefont
  {Huang}}, \bibinfo {author} {\bibfnamefont {D.}~\bibnamefont {Lin}}, \bibinfo
  {author} {\bibfnamefont {C.}~\bibnamefont {Wang}}, \bibinfo {author}
  {\bibfnamefont {W.}~\bibnamefont {Liu}}, \bibinfo {author} {\bibfnamefont
  {S.}~\bibnamefont {Fang}}, \bibinfo {author} {\bibfnamefont {J.}~\bibnamefont
  {Peng}}, \bibinfo {author} {\bibfnamefont {P.}~\bibnamefont {Huang}},\ and\
  \bibinfo {author} {\bibfnamefont {G.}~\bibnamefont {Zeng}},\ }\bibfield
  {title} {\bibinfo {title} {{Continuous-variable quantum key distribution with
  1 Mbps secure key rate}},\ }\href {https://doi.org/10.1364/OE.23.017511}
  {\bibfield  {journal} {\bibinfo  {journal} {Optics Express}\ }\textbf
  {\bibinfo {volume} {23}},\ \bibinfo {pages} {17511} (\bibinfo {year}
  {2015})}\BibitemShut {NoStop}%
\bibitem [{\citenamefont {Tang}\ \emph {et~al.}(2020)\citenamefont {Tang},
  \citenamefont {Kumar}, \citenamefont {Ren}, \citenamefont {Wonfor},
  \citenamefont {Penty},\ and\ \citenamefont {White}}]{Tang2020}%
  \BibitemOpen
  \bibfield  {author} {\bibinfo {author} {\bibfnamefont {X.}~\bibnamefont
  {Tang}}, \bibinfo {author} {\bibfnamefont {R.}~\bibnamefont {Kumar}},
  \bibinfo {author} {\bibfnamefont {S.}~\bibnamefont {Ren}}, \bibinfo {author}
  {\bibfnamefont {A.}~\bibnamefont {Wonfor}}, \bibinfo {author} {\bibfnamefont
  {R.}~\bibnamefont {Penty}},\ and\ \bibinfo {author} {\bibfnamefont
  {I.}~\bibnamefont {White}},\ }\bibfield  {title} {\bibinfo {title}
  {{Performance of continuous variable quantum key distribution system at
  different detector bandwidth}},\ }\href
  {https://doi.org/10.1016/j.optcom.2020.126034} {\bibfield  {journal}
  {\bibinfo  {journal} {Opt. Commun.}\ }\textbf {\bibinfo {volume} {471}},\
  \bibinfo {pages} {126034} (\bibinfo {year} {2020})}\BibitemShut {NoStop}%
\bibitem [{\citenamefont {Sarmiento}\ \emph {et~al.}(2022)\citenamefont
  {Sarmiento}, \citenamefont {Etcheverry}, \citenamefont {Aldama},
  \citenamefont {L{\'{o}}pez}, \citenamefont {Vidarte}, \citenamefont {Xavier},
  \citenamefont {Nolan}, \citenamefont {Stone}, \citenamefont {Li},
  \citenamefont {Loeber},\ and\ \citenamefont {Pruneri}}]{Sarmiento2022}%
  \BibitemOpen
  \bibfield  {author} {\bibinfo {author} {\bibfnamefont {S.}~\bibnamefont
  {Sarmiento}}, \bibinfo {author} {\bibfnamefont {S.}~\bibnamefont
  {Etcheverry}}, \bibinfo {author} {\bibfnamefont {J.}~\bibnamefont {Aldama}},
  \bibinfo {author} {\bibfnamefont {I.~H.}\ \bibnamefont {L{\'{o}}pez}},
  \bibinfo {author} {\bibfnamefont {L.~T.}\ \bibnamefont {Vidarte}}, \bibinfo
  {author} {\bibfnamefont {G.~B.}\ \bibnamefont {Xavier}}, \bibinfo {author}
  {\bibfnamefont {D.~A.}\ \bibnamefont {Nolan}}, \bibinfo {author}
  {\bibfnamefont {J.~S.}\ \bibnamefont {Stone}}, \bibinfo {author}
  {\bibfnamefont {M.~J.}\ \bibnamefont {Li}}, \bibinfo {author} {\bibfnamefont
  {D.}~\bibnamefont {Loeber}},\ and\ \bibinfo {author} {\bibfnamefont
  {V.}~\bibnamefont {Pruneri}},\ }\bibfield  {title} {\bibinfo {title}
  {{Continuous-Variable Quantum Key Distribution over 15 km Multi-Core
  Fiber}},\ }\href {http://arxiv.org/abs/2201.03392} {\ ,\ \bibinfo {pages} {3}
  (\bibinfo {year} {2022})},\ \Eprint {https://arxiv.org/abs/2201.03392}
  {arXiv:2201.03392} \BibitemShut {NoStop}%
\bibitem [{\citenamefont {Lodewyck}\ \emph {et~al.}(2007)\citenamefont
  {Lodewyck}, \citenamefont {Bloch}, \citenamefont {Garc{\'{i}}a-Patr{\'{o}}n},
  \citenamefont {Fossier}, \citenamefont {Karpov}, \citenamefont {Diamanti},
  \citenamefont {Debuisschert}, \citenamefont {Cerf}, \citenamefont
  {Tualle-Brouri}, \citenamefont {McLaughlin},\ and\ \citenamefont
  {Grangier}}]{Lodewyck2007}%
  \BibitemOpen
  \bibfield  {author} {\bibinfo {author} {\bibfnamefont {J.}~\bibnamefont
  {Lodewyck}}, \bibinfo {author} {\bibfnamefont {M.}~\bibnamefont {Bloch}},
  \bibinfo {author} {\bibfnamefont {R.}~\bibnamefont
  {Garc{\'{i}}a-Patr{\'{o}}n}}, \bibinfo {author} {\bibfnamefont
  {S.}~\bibnamefont {Fossier}}, \bibinfo {author} {\bibfnamefont
  {E.}~\bibnamefont {Karpov}}, \bibinfo {author} {\bibfnamefont
  {E.}~\bibnamefont {Diamanti}}, \bibinfo {author} {\bibfnamefont
  {T.}~\bibnamefont {Debuisschert}}, \bibinfo {author} {\bibfnamefont {N.~J.}\
  \bibnamefont {Cerf}}, \bibinfo {author} {\bibfnamefont {R.}~\bibnamefont
  {Tualle-Brouri}}, \bibinfo {author} {\bibfnamefont {S.~W.}\ \bibnamefont
  {McLaughlin}},\ and\ \bibinfo {author} {\bibfnamefont {P.}~\bibnamefont
  {Grangier}},\ }\bibfield  {title} {\bibinfo {title} {{Quantum key
  distribution over 25 km with an all-fiber continuous-variable system}},\
  }\href {https://doi.org/10.1103/PhysRevA.76.042305} {\bibfield  {journal}
  {\bibinfo  {journal} {Phys. Rev. A}\ }\textbf {\bibinfo {volume} {76}},\
  \bibinfo {pages} {042305} (\bibinfo {year} {2007})}\BibitemShut {NoStop}%
\bibitem [{\citenamefont {Shrestha}\ and\ \citenamefont
  {Choi}(2019)}]{Shrestha2019}%
  \BibitemOpen
  \bibfield  {author} {\bibinfo {author} {\bibfnamefont {S.}~\bibnamefont
  {Shrestha}}\ and\ \bibinfo {author} {\bibfnamefont {D.-Y.}\ \bibnamefont
  {Choi}},\ }\bibfield  {title} {\bibinfo {title} {{Rain Attenuation Study over
  an 18 GHz Terrestrial Microwave Link in South Korea}},\ }\href
  {https://doi.org/10.1155/2019/1712791} {\bibfield  {journal} {\bibinfo
  {journal} {Int. J. Antennas Propag}\ }\textbf {\bibinfo {volume} {2019}},\
  \bibinfo {pages} {1} (\bibinfo {year} {2019})}\BibitemShut {NoStop}%
\bibitem [{\citenamefont {Zhao}\ and\ \citenamefont {Wu}(2000)}]{Zhao2000}%
  \BibitemOpen
  \bibfield  {author} {\bibinfo {author} {\bibfnamefont {Z.}~\bibnamefont
  {Zhao}}\ and\ \bibinfo {author} {\bibfnamefont {Z.}~\bibnamefont {Wu}},\
  }\bibfield  {title} {\bibinfo {title} {{Millimeter-wave attenuation due to
  fog and clouds}},\ }\href {https://doi.org/10.1023/A:1006611609450}
  {\bibfield  {journal} {\bibinfo  {journal} {J. Infrared Millim. Terahertz
  Waves}\ }\textbf {\bibinfo {volume} {21}},\ \bibinfo {pages} {1607} (\bibinfo
  {year} {2000})}\BibitemShut {NoStop}%
\bibitem [{\citenamefont {Fi{\v{s}}{\'{a}}k}\ \emph {et~al.}(2006)\citenamefont
  {Fi{\v{s}}{\'{a}}k}, \citenamefont {Řez{\'{a}}{\v{c}}ov{\'{a}}},\ and\
  \citenamefont {Mattanen}}]{Fisak2006}%
  \BibitemOpen
  \bibfield  {author} {\bibinfo {author} {\bibfnamefont {J.}~\bibnamefont
  {Fi{\v{s}}{\'{a}}k}}, \bibinfo {author} {\bibfnamefont {D.}~\bibnamefont
  {Řez{\'{a}}{\v{c}}ov{\'{a}}}},\ and\ \bibinfo {author} {\bibfnamefont
  {J.}~\bibnamefont {Mattanen}},\ }\bibfield  {title} {\bibinfo {title}
  {{Calculated and measured values of liquid water content in clean and
  polluted environments}},\ }\href {https://doi.org/10.1007/s11200-006-0006-z}
  {\bibfield  {journal} {\bibinfo  {journal} {Stud. Geophys. Geod}\ }\textbf
  {\bibinfo {volume} {50}},\ \bibinfo {pages} {121} (\bibinfo {year}
  {2006})}\BibitemShut {NoStop}%
\bibitem [{\citenamefont {Kim}\ \emph {et~al.}(2001)\citenamefont {Kim},
  \citenamefont {McArthur},\ and\ \citenamefont {Korevaar}}]{Kim2001}%
  \BibitemOpen
  \bibfield  {author} {\bibinfo {author} {\bibfnamefont {I.~I.}\ \bibnamefont
  {Kim}}, \bibinfo {author} {\bibfnamefont {B.}~\bibnamefont {McArthur}},\ and\
  \bibinfo {author} {\bibfnamefont {E.~J.}\ \bibnamefont {Korevaar}},\
  }\bibfield  {title} {\bibinfo {title} {{Comparison of laser beam propagation
  at 785 nm and 1550 nm in fog and haze for optical wireless communications}},\
  }in\ \href {https://doi.org/10.1117/12.417512} {\emph {\bibinfo {booktitle}
  {Optical Wireless Communications III}}},\ Vol.\ \bibinfo {volume} {4214}\
  (\bibinfo  {publisher} {SPIE},\ \bibinfo {year} {2001})\BibitemShut {NoStop}%
\bibitem [{\citenamefont {Sanz}\ \emph {et~al.}(2018)\citenamefont {Sanz},
  \citenamefont {Fedorov}, \citenamefont {Deppe},\ and\ \citenamefont
  {Solano}}]{Mikel18}%
  \BibitemOpen
  \bibfield  {author} {\bibinfo {author} {\bibfnamefont {M.}~\bibnamefont
  {Sanz}}, \bibinfo {author} {\bibfnamefont {K.~G.}\ \bibnamefont {Fedorov}},
  \bibinfo {author} {\bibfnamefont {F.}~\bibnamefont {Deppe}},\ and\ \bibinfo
  {author} {\bibfnamefont {E.}~\bibnamefont {Solano}},\ }\bibfield  {title}
  {\bibinfo {title} {Challenges in open-air microwave quantum communication and
  sensing},\ }in\ \href {https://doi.org/10.1109/CAMA.2018.8530599} {\emph
  {\bibinfo {booktitle} {2018 IEEE Conference on Antenna Measurements
  Applications (CAMA)}}}\ (\bibinfo {year} {2018})\BibitemShut {NoStop}%
\end{thebibliography}%

\end{document}